 \documentclass[12pt,preprint,epsfig]{aastex}

 \usepackage[]{graphicx}
 \usepackage{emulateapj5}

\shorttitle{The star formation history of the host galaxies of GRBs}
 \shortauthors{Calura et al.}

\begin{document}


 \title{Constraining the star formation histories of GRB Host Galaxies from their observed abundance patterns}


 \author{Francesco Calura$^{1}$, Miroslava Dessauges-Zavadsky$^2$, Jason X. Prochaska$^3$ and 
 Francesca Matteucci$^{1}$} 
\affil{
1 Dipartimento di Astronomia-Universit\'a di Trieste, Via G. B. Tiepolo
w  11, 34131 Trieste, Italy\\
2 Observatoire de Geneve, Universite de Geneve, 51 Ch. des Maillettes, 1290 Sauverny, Switzerland \\
3 UCO/Lick Observatory, University of California at Santa Cruz, Santa Cruz, CA 95064, USA\\
}
 \email{fcalura@oats.inaf.it}

 %% are identified by the \altaffilmark after each name. Specify alternate

\begin{abstract}
Long-duration Gamma Ray Bursts (GRBs) are linked to the collapse of massive stars and their hosts  
are exclusively identified as active, star forming galaxies. Four long GRBs 
observed at high spectral resolution at redshift $1.5 \le z \le 4$ allowed the determination 
of the elemental abundances for a set of different chemical elements. In this paper, for the first time, 
by means of detailed chemical evolution models taking into account also dust production, we attempt to 
constrain the star formation history of the host galaxies of these GRBs from the study of the 
measured chemical abundances measured in their ISM. We are also able to provide constraints on the age and 
on the dust content of GRB hosts.  
Our results support the hypothesis that long duration GRBs occur preferentially in 
low metallicity, star forming galaxies. 
We compare the  specific star formation rate, namely the star formation rate per unit stellar mass, 
predicted for the hosts of these GRBs  with observational values for GRB hosts distributed across a large redshift range. 
Our models predict a decrease of the specific star formation rate (SSFR) with redshift, consistent with the observed decrease of the comoving 
cosmic SFR density between $z \sim2 $ and $z=0$. 
On the other hand, observed GRB hosts seems to follow an opposite trend in the SSFR vs redshift plot, 
with an increase of the SSFR with decreasing redshift.  Future SSFR determination in larger samples 
of GRB hosts will be important to understand whether this trend is real or due to some selection effect. 
Finally, we compare the SSFR of GRB050730 host with values derived with a sample of 
Quasar damped Lyman alpha systems. Our results indicate that the abundance pattern and the 
specific star formation rates of the host galaxies of these GRBs are basically compatible with the ones determined for a sample of 
Quasar damped Lyman alpha systems, suggesting similar chemical evolution paths. 
\end{abstract} 

 %% and \citet commands to identify citations. The citations are

\section{Introduction}

Gamma-ray bursts (GRBs) are valuable tools to investigate the 
high-redshift Universe. 
The extreme luminosities of their afterglows 
have allowed detailed studies of 
the physical properties of the interstellar medium (ISM) of their host galaxies. 
In particular, in a few cases, 
the determination of their chemical abundance pattern has been possible \citep{SAV03, VRE04, PRO04, PEN06, LEV07, WIE07, PRO07a}, 
as well as the determination of a 
few quantities related to their host galaxies, 
such as their dust content, their stellar mass and 
their star formation rate \citep{BLO98, LEF02, DJO03, CHR04, LEF06, CAS06, SAV08}. \\
\citet{CHE05} first reported on the
chemical abundances for the damped Lyman Alpha system (DLA) associated
with the host galaxy of GRB 050730.  
Their analysis showed this gas was metal poor with
a modest depletion level (see also Starling et al. 2005). 
These results were subsequently
expanded and tabulated by \citet{PRO07a} (hereafter P07, see
also D'Elia et al 2007) and 
we adopt their values adjusted to the solar relative abundances of \citet{GRE07}. \\
In this paper, for the first time we aim at determining the star formation history 
of some GRB DLA systems by analysing  the chemical abundances measured 
in their afterglow spectra. For this purpose, we use a detailed chemical evolution 
model,  which allows us to predict the evolution of the abundances of   
the chemical elements studied in the GRB DLAs of P07.   
Our aim is to constrain several relevant astrophysical quantities, 
such as the star formation 
rates, the dust content and the ages of the host galaxies of these GRBs. 
Our approach has already turned out to be fruitful to study Quasar DLAs  
on the basis of their chemical abundance pattern, allowing us to derive 
crucial information on their nature and on their physical 
properties \citep{DES04, DES07}. 
The plan of this paper is as follows. 
In Section 2, we briefly introduce the chemical evolution model. 
In Section 3, we present our results and in Section 4 we draw some conclusions.

\section{The chemical evolution model}
\subsection{The dwarf irregular model}
The model used to derive the star formation histories of  these GRB DLAs host galaxies 
is similar to that 
developed by \citet{BRA98} 
to study dwarf irregular galaxies. 
Since some chemical species studied in this paper are refractory (C, O, Si, Mg, Fe, Ni),  
in the chemical evolution model used here 
we include also a detailed treatment of dust production and destruction following the work of \citet{CAL08}(CPM08, see section 2.2). \\
We choose to use the model for dwarf irregular galaxies since several 
observational investigations  
provided strong evidence that most of the GRBs originate 
in gas rich, star-forming sub-luminous ($L<L_{*}$) galaxies with relatively low 
metallicities ($Z < Z_{\odot}$) \citep{BLO98, FRU99, PRO04, VRE04}. \\
%Furthermore, in Sect. 3.1, 
%we will clearly show how chemical evolution models for ellipticals and spiral galaxies 
%are inadequate to describe the abundance pattern observed in the GRB 050730 host galaxy. \\
The dwarf galaxy is assumed to form by means 
of a continuous infall of 
pristine gas until a mass $M_{tot}$ is accumulated.  
The evolution of dwarf irregular galaxies is characterized  
by a continuous star formation history.
% characterized by a low star formation efficiency 
%($\nu \le 0.5 Gyr^{-1}$). 
%parameters which need to be constrained 
%are the number of bursts of star formation and 
%for each burst the efficiency, 
%$\nu$, and the burst duration, $\Delta t$ 
%(Dessages-Zavadsky et al. 04, 06, 07). 
%Since there is a large number of evidences 
%which favour the link between long GRBs and the explosion of massive stars, occurring 
%in active star forming regions, for our model we assume 
%a continuous star formation history. 
%Previous works on the chemical evolution of damped Lyman Alpha 
%systems (Dessages-Zavadsky et al. 04, 07) showed 
%that the abundance patterns of a system 
%whose SF proceeded 
%with short bursts of high efficiency and  
%by means of a low-efficiency and prolonged SF history  
%were practically equivalent.   
%In this paper, for our model we assume
%a continuous star formation history, i.e. one single burst 
%of very long duration. 
%This assumption allows us to get rid 
%of the parameter $\Delta t$ and of the number of bursts, 
%with  the star formation efficiency 
%$\nu$ the only parameter left to test.  \\
The star 
formation rate $\psi(t)$ is directly proportional the gas 
fraction $G(t)$ at the time $t$, according to the Schmidt law \citep{SCH59}
\begin{equation}
\psi(t)=\nu G(t).
\end{equation}
where $\nu$ is the efficiency of star formation.  
We take into account the energy feedback from both Type Ia and 
Type II supernovae (SNe). 
They  are responsible for the onset of 
a galactic wind, when the thermal 
energy of the ISM exceeds its binding energy, 
which is related to the presence of a dark matter 
halo in which the galaxy is embedded (for more details, see Bradamante et al. 1998, Lanfranchi \& Matteucci 2003). 
The thermal energy of the gas is controlled mainly by the 
thermalization efficiencies of the supernovae explosions
($\eta_{SNII}$ for SNe II and $\eta_{SNIa}$ for SNeIa) and stellar
winds ($\eta_{SW}$), which control the fraction of the energy restored to the ISM which is 
converted to thermal energy of the gas (see Bradamante
et al. 1998). The binding energy of the gas, on the other hand, is strongly
influenced by  assumptions concerning the presence and distribution
of dark matter (Matteucci 1992). A diffuse ($R_e/R_d$=0.1,  where
$R_e$ is the effective radius of the galaxy and $R_d$ is  the radius
of the dark matter core) but relatively massive  ($M_{dark}/M_{Lum}=10$) dark
halo has been assumed for each galaxy. 
The time at which the wind develops depends on the assumed star formation efficiency. 
In general, the higher the SF efficiency, the earlier the wind develops. \\
%The models used throughout this paper are characterized by very low star formation efficiencies 
%and by young ages, much lower than the times of occurrence of the galactic winds. 
%For these reasons, the galactic winds have no effect on the main results obtained in this paper. \\
No instantaneous recycling approximation is adopted, i.e. the stellar lifetimes are 
properly taken into account. This allows us to compute in detail 
the contributions by low mass stars, Type Ia, 
Type II SNe to the chemical enrichment of the ISM. 
The main physical quantities related to these processes are the 
stellar yields, representing the fractions of processed matter 
restored by the stars into the ISM. 
The stellar yields used in this work are from \citet{MEY02} for 
low and intermediate mass stars (LIMS), from \citet{WOO95} for massive stars 
and from \citet{IWA99} for Type Ia SNe.  
%For a detailed description of the chemical evolution equation and 
%of how low and intermediate mass stars (LIMS), type Ia, and type II SNe  contribute to the chemical enrichment of the ISM, 
%see \citet{MAT86}.  
We assume a Salpeter initial mass function (IMF). With such an IMF, it is possible 
to reproduce the chemical abundances and the observational features of  
dwarf irregulars \citep{CAL06}  and to account for the local metal budget \citep{CAL04}.
We assume a cosmological model characterized by $\Omega_m=0.3$, 
$\Omega_{\Lambda}=0.7$ and a Hubble constant $H_{0}=70 km s^{-1} Mpc^{-1}$.   

\subsection{N nucleosynthesis: indications from QSO DLAs}
\label{Nprod}
\begin{figure*}[t]
\includegraphics[height=30pc,width=30pc]{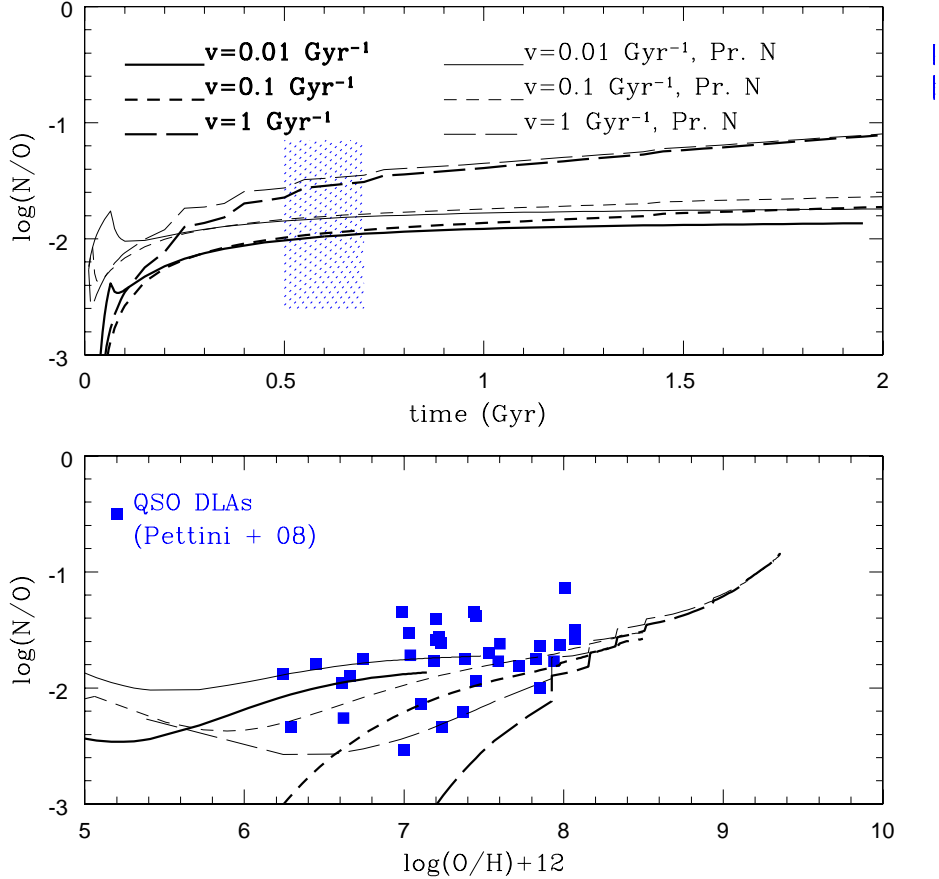}
\caption{ \emph{Upper panel}
: Predicted log(O/N) vs time for models adopting various star formation efficiencies. 
The thin lines include primary N in massive stars following Matteucci (1986). 
The thick lines do not include primary N production in massive stars. The shaded area represents the range of log(N/O) 
values observed by various authors (Pettini et al. 2008 and references therein) in a sample of QSO Damped Lyman Alpha systems.
\emph{Lower panel}: Predicted log(O/N) vs log(O/H)+12. Thick and thin lines as in the upper panel. 
The solid squares are observational values in high-redshift QSO DLAs (Pettini et al. 2008 and references therein). }
\label{fig1}
\end{figure*}

Although N nucleosynthesis is still a matter of debate, most of N production 
arise from low and intermediate mass stars since
massive stars produce only a small fraction of it 
(e.g. Chiappini, Matteucci \& Ballero 2005). In particular, in the solar neighborhood the typical timescale for 
N enrichment should be 0.25-0.30 Gyr. However,  the N produced in massive stars 
is important for understanding N abundances at very low metallicities. 
In principle, N produced in massive stars should be secondary, 
namely produced by means of the CNO cycle starting from the C and O abundances 
originally present in the star. In the last years Meynet \& Maeder (2000, 2002) 
and more recently Chiappini et al (2006) have shown that 
very metal poor stars can be fast rotators and that rotation can produce primary N, 
namely N originating from the C and O produced by the star, not those originally present in the star at birth. 
In this case, a high N  abundance is expected at very low metallicities. 
It is not clear whether this high production of primary N in massive stars 
continue at intermediate metallicities.
In our case we adopt the standard prescriptions 
for secondary N in massive stars and secondary plus primary N in low and intermediate mass stars. \\
In Fig.~\ref{fig1}, we show the predicted evolution of log(N/O) vs time (upper panel) and vs metallicity (lower panel) assuming 
various star formation efficiencies. The models are 
computed with the nucleosynthesis prescriptions used here (thick lines) and assuming primary N in massive stars (thin lines), following 
Matteucci (1986). 
The predictions are compared with the observed N/O values measured in high redshift QSO DLAs by various authors 
(Pettini et al. 2008 and references therein). 
The prescriptions adopted here are in good agreement with the most recent data published by Pettini et al. (2008) for DLAs , 
whose metallicities are not as low as metal poor halo stars where a high N abundance seems to be found (Spite et al. 2005). 
%On the other hand, the assumption of primary N in massive stars predicts too high N/O values immediately after 
%the beginning of star formation. \\
The assumption of primary N in massive stars predicts high N/O values at low metallicity, i.e. 
immediately after the beginning of star formation. However, as can be seen from 
Fig.~\ref{fig1}, the highest N/O values in DLAs can be explained also without assuming 
primary N production in massive stars, but by assuming a low star formation efficiency.\\
Without primary N production in massive stars, the value log(N/O)=-2.6, corresponding to the lowest value observed in DLAs, is reached by our models 
after 0.05 Gyr with $\nu=0.01$ Gyr$^{-1}$ and after 0.1 Gyr with $\nu=1$ Gyr$^{-1}$. 
Finally, it is worth stressing that, as remarked also by Chiappini et al. (2003), the stellar yields of  Meynet \& Maeder (2002) 
do not include the third dredge-up and hot-bottom burning,  thus their N yields for intermediate mass 
stars should be regarded as lower limits.

\subsection{Dust production and destruction} 
The model for dust evolution used in this paper adopts  
the same prescriptions as CPM08. For a detailed description of the dust evolution model, 
we address the reader to that paper. Here, we briefly summarize our assumptions.\\
For the refractory chemical element labeled $i$, 
a fraction $\delta^{SW}_{i}$, $\delta^{Ia}_{i}$, and  $\delta^{II}_{i}$ is incorporated in dust grains by 
low and intermediate mass stars, type Ia SNe, and type II SNe, respectively. These quantities are the dust condensation efficiencies 
 of the element $i$ in various stellar objects. 
Here we assume $\delta^{SW}_{i}=\delta^{Ia}_{i}=\delta^{II}_{i} \equiv \delta_{i} =0.1$. 
This choice is supported by recent mid-infrared observations of one Supernova 
(Sugerman et al. 2006), which provided an upper limit of $\delta^{II}_{i} \le 0.12$. 
The value assumed here is also supported by  theoretical studies of the local dust cycle 
(Edmunds 2001).  
Our assumption is compatible with values provided by other theoretical and observational studies (see Morgan \& Edmunds 2003). 
However, other studies suggest that the dust condensation efficiencies of SNe may be even lower, with values ranging from 
$\sim 2 \times 10^{-4}$ to 0.15 (Zhukovska et al. 2008).\\ 
%We constrain these parameters by  requiring that 
%the depletion pattern of the model describing the GRB 
%host galaxy is compatible with the ones observed in the Milky Way or in the 
%Magellanic Clouds. \\
CPM08 showed that the dust depletion pattern is strongly dependent  on the star formation history. 
The depletion pattern observed in galaxies of different morphological types 
provides us with the most complete set of dust depletion data for a range of different star formation histories. \\
The refractory elements involved in this study are C and O which in the local ISM show mild depletion levels, and 
Mg, Si, Ni, and Fe, which locally show higher depletion levels \citep{SAV96}. 
A debate is currently open on the nature of S and on whether it may be 
incorporated into dust grains or not. 
Observational studies of the depletion patterns of the cold ISM of the Milky Way 
report for S depletion values between 0 and -0.2 dex \citep{WEL99a}.  
More recent data, based on observational determinations of the gas-phase abundances of 
some chemical elements in the Local Interstellar Cloud, support a 
considerable depletion for S of -0.3 dex \citep{KIM03}. 
However, these high depletion values have been derived by assuming as cosmic S abundance 
the solar value observed by Grevesse \& Sauval (1998), i.e. $A_{S}=21.4$ per $10^6$ hydrogen atoms. 
The recent update by Grevesse et al. (2007) suggests a solar S abundance $A_{S}=14$ per $10^6$ Hydrogen Atoms, 
which implies a S dust fraction in the Local Interstellar Cloud lower than 0.4, i.e. a logarithmic depletion higher than -0.2 dex.  
However, other indirect arguments indicate that S may have a refractory nature. 
Detailed models of dark insterstellar clouds chemistry suggest that 
sulphur must be depleted by two to three orders of magnitude in dark regions \citep{RUF99, SCA03}. 
A recent  investigation of dust and element abundances in planetary nebulae has shown that in these systems, 
depletion is likely to be severe for various species, including S \citep{PHI07}. 
In cometary dust particles, S is present in iron sulphide (FeS) grains \citep{LOD03}. 
A broad feature ascribed to FeS emission has been recently seen in the infrared spectra of young stellar objects \citep{KELL02}. 
Furthermore, an indirect evidence of the fact that S may suffer depletion comes from the fact that its condensation temperature 
is considerably higher than the ones of C and O, which are known to be depleted in the gas phase 
(see Welty et al. 1999). 
From a photospheric abundance study of a RV Tauri star in the Large Magellanic Cloud, 
Reyniers \&  Van Winckel (2007) showed  how the  element abundance  anti-correlates 
with the condensation temperature, possibly implying a correlation between the depletion and the condensation temperature. 
In the light of these facts, in our study we consider two different scenarios 
for dust grain composition. 
In the ``expected dust'' scenario, we assume that the only elements going into dust grains 
are Ni and Fe. 
The reason of the name chosen for this scenario  lies in the fact that mainly indirect evidences 
indicate that S could 
be a refractory element. 
We consider also an ``alternative dust'' scenario, where also S is refractory.\\
Dust grains can be destroyed by the propagation 
of supernova shock waves in the warm/ionised interstellar medium \citep{MCK89, JON94}. 
If $G_{dust,i}$ is the fraction of the element $i$ locked into dust and $G$ is the gas fraction, 
the destruction rate is calculated as 
$G_{dust,i}/\tau_{destr}$, where $\tau_{destr}$ is the dust destruction timescale, calculated as: 
\begin{equation}
\tau_{destr}=(\epsilon M_{SNR})^{-1} \frac{G}{R_{SN}}. 
\end{equation} 
\citep{MCK89, DWE98}. 
$M_{SNR}=1300 M_{\odot}$ is the mass of the interstellar gas swept up by the SN remnant \citep{MCK89, DWE07}. 
$R_{SN}$ is the total SNe rate, i.e. the sum of the rates of Type Ia and Type II SNe.\\ 
Unless otherwise specified, 
we assume that no dust accretion is taking place in the GRB host galaxies. This phenomenon occurs in $H_{2}$-rich 
molecular clouds \citep{DWE98, CAL08}. 
Our choice is motivated by the fact that 
very little molecular H is observed in local dwarfs, with molecular-to-atomic gas fractions 
of $\sim 10 \%$ or lower \citep{LIS98, CLA96}. Our assumption is  
further supported by the fact that a very little amount of $H_{2}$ observed in the spectra of 
GRB afterglows \citep{VRE04, FYN06, TUM07, WHA08}.  
This might indicate that in the presence of intense SF, molecular clouds could rapidly dissolve, 
likely allowing very little dust accretion to occur. We will test the reliability of this 
assumption while discussing our results. \\
%%%%%%%%%%%%%%%%%%%%%%%%%%%%%%%%%%%%%%%%%%%%%%%%%%%%%%%%%%%%%%%%%%%%%%%%%%%%%%%%%%%%%%%%%%%%%%%%%%%%%%%%%%%%%%%%%%%
%%%%%%%%%%%%%%%%%%%%%%%%%%%%%%%%%%%%%%%%%%%%%%%%%%%%%%%%%%%%%%%%%%%%%%%%%%%%%%%%%%%%%%%%%%%%%%%%%%%%%%%%%%%%%%%%%%%
\begin{figure*}
\includegraphics[width=\columnwidth]{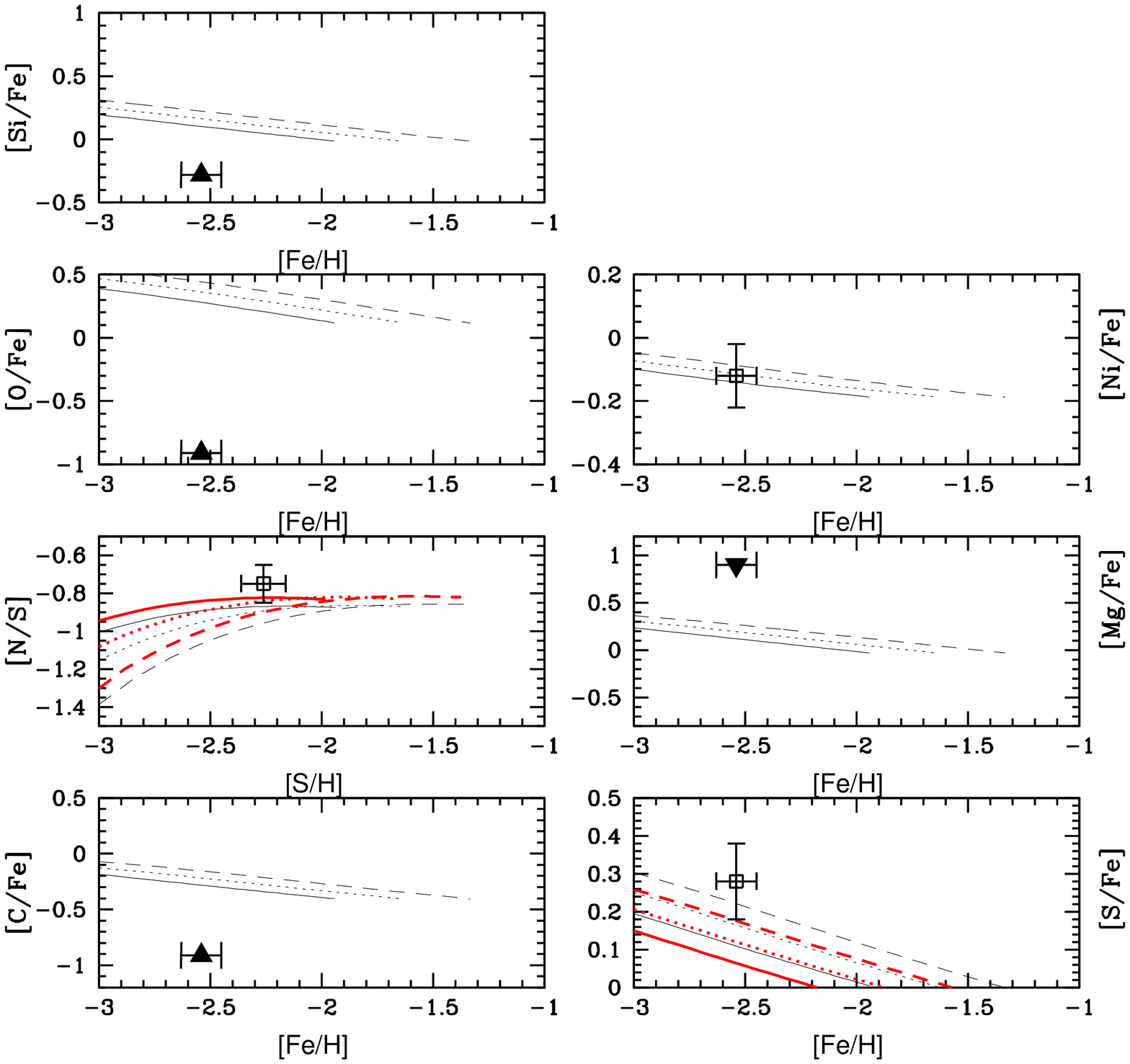}
\caption{ Observed abundance ratios versus metallicity for the host galaxy of GRB 050730   
as derived by \citet{PRO07a} (empty squares with error bars and solid triangles). 
The thin solid line, dotted line and dashed line 
are predictions computed by means of a chemical evolution model for 
a dwarf irregular galaxy assuming for the star formation efficiency  
$\nu=0.005 Gyr^{-1}$, $\nu=0.01 Gyr^{-1}$, and $\nu=0.02 Gyr^{-1}$, respectively, with ``expected dust'' composition. 
The thick lines are as above, but with ``alternative dust'' composition, i.e. with S depleted. }
\label{fig2}
\end{figure*}
%%%%%%%%%%%%%%%%%%%%%%%%%%%%%%%%%%%%%%%%%%%%%%%%%%%%%%%%%%%%%%%%%%%%%%%%%%%%%%%%%%%%%%%%%%%%%%%%%%%%%%%%%%%%%%%%%%
\begin{figure*}
\includegraphics[width=\columnwidth]{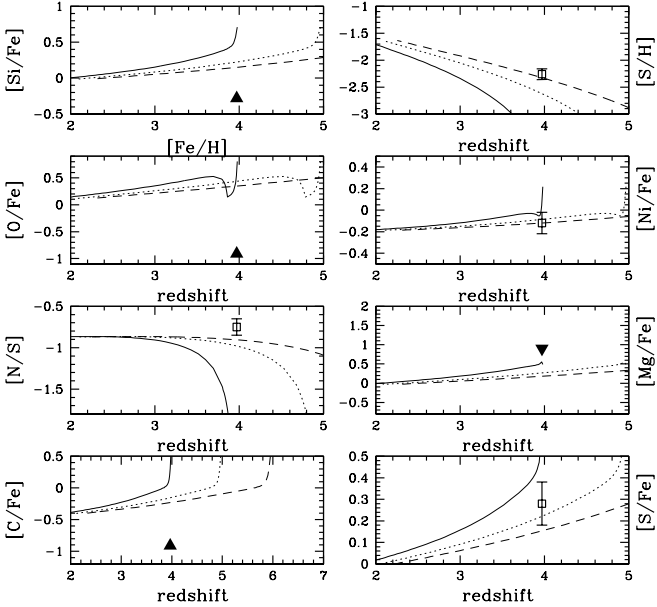}
\caption{ Observed abundance ratios versus redshift for the host galaxy of GRB 050730   
as derived by \citet{PRO07a} (empty squares with error bars and solid triangles).  
The thin solid line, dotted line and dashed line 
are predictions computed by means of the best model 
($\nu=0.01 Gyr^{-1}$, ``expected dust'' composition) and assuming $z_f=4$, $z_f=5$ and $z_f=6$, 
respectively. }
\label{fig3}
\end{figure*}
%%%%%%%%%%%%%%%%%%%%%%%%%%%%%%%%%%%%%%%%%%%%%%%%%%%%%%%%%%%%%%%%%%%%%%%%%%%%%%%%%%%%%%%%%%%%%%%%%%%%%%%%%%%%%%%%%%
%\begin{table}
%\caption[]{Main Model parameters}
%\begin{tabular}{l|c|c}
%\\[-2.0ex] 
%\hline
%\\[-2.5ex]
% Parameter   & Value adopted         &   Constrained by                   \\
%\hline
%IMF         & Salpeter               & Local metal budget (Calura \& Matteucci 2004) \\
% \ny        & free                   &          \\
%\hline
%$\delta_i$    &    0.1                 & SN ejecta (Sugerman et al. 2006)      \\
%$M_{SNR}$  &    $1300 M_{\odot}$            &   Estimate for a 3-phases ISM, (McKee 1989)  \\
%\hline
%\hline
%\end{tabular}
%\label{tab1}
%\end{table}
%%%%%%%%%%%%%%%%%%%%%%%%%%%%%%%%%%%%%%%%%%%%%%%%%%%%%%%%%%%%%%%%%%%%%%%%%%%%%%%%%%
%%%%%%%%%%%%%%%%%%%%%%%%%%%%%%%%%%%%%%%%%%%%%%%%%%%%%%%%%%%%%%%%%%%%%%%%%%%%%%%%%%
\renewcommand{\baselinestretch}{1.0}
\begin{table*}
\centering
\caption{}
\begin{tabular}{lcccccccccccccc}
\\[-2.0ex] 
\hline
\\[-2.5ex]
\multicolumn{1}{l}{Ab. ratio}&\multicolumn{2}{c}{GRB 050730}&\multicolumn{2}{c}{GRB 050820}&\multicolumn{2}{c}{GRB 051111}&\multicolumn{2}{c}{GRB 060418}\\%&\multicolumn{2}{c}{Metallicity}\\
%\multicolumn{1}{l}{}&\multicolumn{2}{c}{(10$^{10}$L$_{\odot}$) }&\multicolumn{2}{c}{(10$^{8}$M$_{\odot}$)}&\multicolumn{2}{c}{(M$_{\odot}$/L$_{\odot}$)}\\%&\multicolumn{2}{c}{}&\multicolumn{2}{c}{}\\
\hline 
\multicolumn{1}{c}{}&\multicolumn{1}{c}{Obs}&\multicolumn{1}{c}{Pred}&\multicolumn{1}{c}{Obs}&\multicolumn{1}{c}{Pred}&\multicolumn{1}{c}{Obs}&\multicolumn{1}{c}{Pred}&\multicolumn{1}{c}{Obs}&\multicolumn{1}{c}{Pred}\\%&\multicolumn{1}{c}{Obs}&\multicolumn{1}{c}{Pred}\\
\hline 
\hline
\\[-1.0ex]
\multicolumn{1}{l}{[S/H]}&\multicolumn{1}{c}{-2.26}&\multicolumn{1}{c}{-2.34}&\multicolumn{1}{c}{-0.63}&\multicolumn{1}{c}{-0.76}&
\multicolumn{1}{c}{....}&\multicolumn{1}{c}{....}&\multicolumn{1}{c}{....} &\multicolumn{1}{c}{-1.27} \\

%\multicolumn{1}{l}{[Si/H]}&\multicolumn{1}{c}{$>$-2.82}&\multicolumn{1}{c}{-2.34}&\multicolumn{1}{c}{$>$-1.19}&\multicolumn{1}{c}{-1.82}&
%\multicolumn{1}{c}{$>-1.06$}&\multicolumn{1}{c}{....}&\multicolumn{1}{c}{$>$-1.65} &\multicolumn{1}{c}{-2.13} \\

\multicolumn{1}{l}{[Fe/H]}&\multicolumn{1}{c}{-2.54}&\multicolumn{1}{c}{-2.50}&\multicolumn{1}{c}{-1.64}&\multicolumn{1}{c}{-1.81}&
\multicolumn{1}{c}{-1.78}&\multicolumn{1}{c}{....}&\multicolumn{1}{c}{-2.26} &\multicolumn{1}{c}{-2.22} \\

%\multicolumn{1}{l}{[Zn/H]}&\multicolumn{1}{c}{....}&\multicolumn{1}{c}{---}&\multicolumn{1}{c}{....-----}&\multicolumn{1}{c}{----}&
%\multicolumn{1}{c}{........---}&\multicolumn{1}{c}{....}&\multicolumn{1}{c}{........-} &\multicolumn{1}{c}{----} \\

\multicolumn{1}{l}{[C/Fe]}&\multicolumn{1}{c}{$>$-0.91}&\multicolumn{1}{c}{-0.23}&\multicolumn{1}{c}{$>$-0.34}&\multicolumn{1}{c}{-0.42}&
\multicolumn{1}{c}{....}&\multicolumn{1}{c}{....}&\multicolumn{1}{c}{....} &\multicolumn{1}{c}{-0.32} \\

\multicolumn{1}{l}{[N/S]}&\multicolumn{1}{c}{-0.75}&\multicolumn{1}{c}{-0.91}&\multicolumn{1}{c}{$>$-0.57}&\multicolumn{1}{c}{-0.79}&
\multicolumn{1}{c}{....}&\multicolumn{1}{c}{....}&\multicolumn{1}{c}{....} &\multicolumn{1}{c}{-0.88} \\

\multicolumn{1}{l}{[S/Fe]}&\multicolumn{1}{c}{+0.28}&\multicolumn{1}{c}{ 0.16}&\multicolumn{1}{c}{+1.01}&\multicolumn{1}{c}{1.05}&
\multicolumn{1}{c}{....}&\multicolumn{1}{c}{....}&\multicolumn{1}{c}{....} &\multicolumn{1}{c}{0.93} \\

\multicolumn{1}{l}{[Si/Fe]}&\multicolumn{1}{c}{$>$-0.28}&\multicolumn{1}{c}{0.15}&\multicolumn{1}{c}{$>$+0.45}&\multicolumn{1}{c}{-0.01}&
\multicolumn{1}{c}{$>$+0.72}&\multicolumn{1}{c}{....}&\multicolumn{1}{c}{$>$+0.61} &\multicolumn{1}{c}{0.09} \\

\multicolumn{1}{l}{[O/Fe]}&\multicolumn{1}{c}{$>$-0.91}&\multicolumn{1}{c}{0.35}&\multicolumn{1}{c}{$>$-0.32}&\multicolumn{1}{c}{0.07}&
\multicolumn{1}{c}{....}&\multicolumn{1}{c}{....}&\multicolumn{1}{c}{....} &\multicolumn{1}{c}{0.34} \\

\multicolumn{1}{l}{[Mg/Fe]}&\multicolumn{1}{c}{$<$+0.90}&\multicolumn{1}{c}{0.18}&\multicolumn{1}{c}{+0.93}&\multicolumn{1}{c}{-0.10}&
\multicolumn{1}{c}{$>$-0.98}&\multicolumn{1}{c}{....}&\multicolumn{1}{c}{$>$-0.97} &\multicolumn{1}{c}{0.03} \\

\multicolumn{1}{l}{[Ni/Fe]}&\multicolumn{1}{c}{-0.12}&\multicolumn{1}{c}{-0.12}&\multicolumn{1}{c}{+0.10}&\multicolumn{1}{c}{-0.19}&
\multicolumn{1}{c}{-0.09}&\multicolumn{1}{c}{....}&\multicolumn{1}{c}{-0.11} &\multicolumn{1}{c}{-0.16} \\

\multicolumn{1}{l}{[Zn/Fe]}&\multicolumn{1}{c}{....}&\multicolumn{1}{c}{-0.05}&\multicolumn{1}{c}{+0.95}&\multicolumn{1}{c}{0.94}&
\multicolumn{1}{c}{$>$+1.14}&\multicolumn{1}{c}{....}&\multicolumn{1}{c}{+0.65} &\multicolumn{1}{c}{0.74} \\

\\[-1.0ex]
\hline
\hline
\end{tabular}
\flushleft
\small
Abundance ratios used in this work as measured in four GRB host galaxies and as predicted by means of our best models. 
All the abundances are normalized to the solar values, from Grevesse et al. (2007).
%Observed and predicted present-day properties of the galactic morphological types studied in this work. 
%The galactic Hubble types  are listed in column 1. In  
%columns 2, 3, 4, 5 and 6  we present the observed and predicted values for the blue luminosity, the HI mass, the HI mass to light ratio, 
%the B-V color and the metallicity, respectively. 
%\emph{References}: $^1$Sansom et al. (2000); $^2$Robert \& Haynes 1994, $^3$Garland et al. (2004); $^4$Hunter \& Elmegreen (2004); $^{5}$Kobayashi \& Arimoto (1999); $^{6}$Vila-Costas \& Edmunds (1992). \\
%\emph{Notes}:$^a$For E/S0 galaxies, the chemical abundances reported in the table are the stellar ones. For S0a/b, Sbc/d and Irr galaxies, the chemical 
%abundances reported in the table are the ones measured in H$_{II}$ regions.
\label{properties}
\end{table*}
%%%%%%%%%%%%%%%%%%%%%%%%%%%%%%%%%%%%%%%%%%%%%%%

\section{Results}
In this section, we attempt to constrain the star formation histories for 
the set of 4 GRB host galaxies. The abundance ratios measured by  P07
for the 4 GRB hosts are presented in Table 1, as well as the predicted abundances  
computed by means of the best model, when possible. \\
We address the reader to the paper by \citet{PRO07a} for further 
details on the GRB observations and on the techniques used to derive 
the ISM abundances of the GRB host galaxies. 

\subsection{Abundance ratios as a diagnostic to infer the star formation history}
In chemical evolution models, the abundance ratios between two 
elements formed on 
different timescales can be used as "cosmic clocks" and provide us with information on the roles of 
LIMS and SNe in the enrichment process \citep{MAT01}. 
In particular, the study of abundance ratios such 
as [$\alpha$/Fe] and [N/$\alpha$]
\footnote{All the abundances between two different elements X and Y 
are expressed as $[X/Y]=log(X/Y)-log(X/Y)_{\odot}$, where  $(X/Y)$ and $(X/Y)_\odot$ are 
the ratios between the mass fractions of X and Y in the ISM and in the sun, respectively. 
We use the set of solar abundances as determined by \citet{GRE07}.} is quite useful, since the $\alpha$-elements (O, S, Mg) 
are produced on short timescales by Type II 
SNe, whereas the Fe-peak elements and nitrogen are produced on long 
timescales by Type Ia SNe and low and intermediate-mass stars, respectively. 
\citet{CAL03b, DES04,DES07} have shown that 
the simultaneous study of the abundance ratios between different elements as functions 
of a metallicity tracer (such as [S/H] or [Fe/H] ) can be used to constrain the nature and 
the star formation history 
of a given system, 
 whereas the study of the abundance ratios versus redshift can be used 
to derive constraints on the age of the system. \\
In these 4 GRB DLAs, P07  have determined  
the abundance ratios for various elements, with typical errors of $\sim$0.1 dex. 
For several other elements, only upper or lower limits were derived for the abundance ratios (see Tab. 1). 
We use the complete set of elements studied by P07 in order to constrain the SF history of the host galaxies of these 4 GRBs. 
In the following, we will present 
the results of our analysis, case by case.

\subsection{GRB 050730}
\label{050730}
%\subsection{Star formation efficiency and age}
%For our analysis, we will focus on the abundance ratios [N/S], [Ni/Fe] and [S/Fe]. 
%The star formation history of  the host galaxy of GRB 050730 is determined by the star 
%formation efficiency of the 
In this Section, our aim is to  constrain 
the main parameter of the star formation history of  the host galaxy of GRB 050730, i.e. its star formation 
efficiency $\nu$. We need to find the model which reproduces at best the 
observed abundance ratios. \\
In this study, the only parameter allowed to vary is the star formation efficiency. Its value is set 
by requiring that, with the adopted SF efficiency, the observed abundance ratios are reproduced.\\
We assume various values for the star formation efficiency $\nu$, spanning from 
$\nu_{min}=0.005 Gyr^-1$ to $\nu_{max}=0.1 Gyr^-1$.  
%For each model and for each chemical element, we vary  
%the dust condensation efficiencies $\delta_i$ in the range 
%$0.3 \le \delta_i \le 1$,
%i.e. between the two most extreme values of the  dust fractions 
% observed in the MW and in the LMC/SMC.
%We are confident that this method will allow us to constrain 
%robustly and univoquely  the range of models which reproduce at best the observed values. 
%We have checked that by increasing the SF efficiency 
%to values higher than $\nu=0.03$, within the ``expected dust'' scenario  
%the predicted abundance ratios 
%are compatible with the measures reported in Fig. 2 only for values of $f_i$ lower than 
%dust fractions observed in the Milky Way and in the Magellanic Clouds. 
In figure 2, we show the predicted evolution of various abundance ratios vs metallicity, 
traced either by [S/H] or by [Fe/H], calculated for a few models assuming different  SF efficiency values 
and compared with the observed abundance ratios.\\
Any model with a star formation efficiency $\nu \le 0.005 Gyr^{-1}$ is rejected, 
since they match the abundances of the 
non-refractory element $N$ at ages larger than 1.53 Gyr, which is 
the age of the Universe 
corresponding to the redshift of the GRB $z_{GRB}=3.96$. 
In Fig. 2, the thin lines represent the predictions computed with 
the ``expected dust'' scenario, i.e. where S is undepleted whereas  the thick lines 
are the predictions within the ``alternative dust'' scenario, where S is depleted. \\
Some of the shown observational  abundance ratios are lower limits ([C/Fe, O/Fe] or 
upper limits [Mg/Fe], and do not provide any strong constraint which can help us decide 
which model is the best in reproducing the data. 
Useful information come from the [N/S] vs [S/H] and [S/Fe] vs 
[Fe/H] plots. The [N/S] vs [S/H] plot tells us that, without taking into account S depletion, 
no model allows us to reproduce the observed [N/S] value. On the other hand, 
a small S depletion of $\sim0.1$ dex is enough to have the observed [N/S] reproduced by 
the two models with $\nu=0.005$ Gyr$^{-1}$ 
and with $\nu=0.01$ Gyr$^{-1}$. 
However, as explained in Sect.~\ref{Nprod}, our predicted 
[N/S] should be regarded as a lower limit, since the yields of Meynet \& Maeder (2002) 
for low and intermediate mass stars neglect the third dredge up, which may play 
an important role in N production (van den Hoek \& Groenewegen 1997). \\
%The effects of varying the N yields are investigated in sect. ---. 
%For purposes of homogeneity, here we prefer to use the massive stars yields by WW95, since  several 
%elements produced by massive stars studied in this paper (S, Mg) have not been 
%investigated in the paper by Meynet \& Maeder (2002). 
%The use of MM02 yields for rotating massive stars of solar metallicity 
%has the effect of increasing the [N/H] abundance by no more than 0.1 dex. \\
%Hence, we can regard this value as an upper limit for N production by massive stars. \\
The [S/Fe] vs [Fe/H] plot indicates that with S undepleted, 
the two models with $\nu=0.01$ Gyr$^{-1}$ 
and $\nu=0.02$ Gyr$^{-1}$ reproduce the observed [S/Fe], however, once we take into account S depletion, 
only the model with $\nu=0.02$ Gyr$^{-1}$ is consistent with the data. 
On the other hand, the [Ni/Fe] vs [Fe/H] plot does not allow us to disentangle among the different models, since models 
with various SF efficiencies present very similar [Ni/Fe] values. \\
From the present study, we conclude that, within the ``expected dust'' scenario, i.e. neglecting any S depletion, 
the best model is the one with $\nu=0.01$ Gyr$^{-1}$, 
with the small discrepancy between our predictions and the models possibly due to the fact that we are underestimating  
[N/H] since we assume no N production in massive stars.  \\
Within the ``alternative dust'' scenario, the best model is the one with $\nu=0.02$  Gyr$^{-1}$. Any model with SF efficiency $\nu=0.03$ Gyr$^{-1}$ 
or higher  
leads us to underestimate the [N/S] value by $\sim 0.2$ dex. This can maybe reconciled 
by assuming primary N production in massive stars and more accurate yields for low and intermediate mass stars, 
including also the third dredge up. 
However, since a further investigation of the N is beyond the aim of the present paper, 
we think it is reasonable to reject any model underestimating [N/S] by more than 0.2 dex ,
corresponding to the total error bar of the measured abundances considered here. \\
Our conclusion is that, 
within the ``expected dust scenario'', the best model is the one with with $\nu=0.01$ Gyr$^{-1}$, whereas 
within the `alternative dust'' scenario, the best model is the one with $\nu=0.02$. \\
%Being the diffence between these two models 
%very small and since the effects of dust depletion are very small in both cases, in the following we will use the model 
%with $\nu=0.01$ Gyr$^{-1}$ computed with the  ``expected dust'' composition to study the age and the SFR of 
%the host galaxy if GRB 050730. 
%the best compromise between the various abundance ratios is obtained by the model with $\nu=0.01$ Gyr$^{-1}$, 
%within the ``expected dust scenario'' (i.e. with S undepleted).  
Once we have constrained the SF efficiency of the GRB 050730 host galaxy, 
by studying its abundance ratios vs redshift we can derive constraints on its age. 
In the following, we will use the two best models of the ``expected dust scenario'' and  ``alternative dust scenario'' 
to derive a value for  the age of the host galaxy of the GRB. \\
In Fig.~\ref{fig3}, we show the predicted redshift evolution of various abundance ratios, 
compared with the abundance ratios observed for GRB 050730. 
We show the predictions computed by means of the best model of the ``expected dust `` scenario by   
assuming three different values for the 
redshift of formation $z_f$, defined as the redshift at which the star formation turns on 
in the GRB host galaxy. 
The GRB host abundances are best reproduced by assuming $z_{f} = 6$. 
The difference between $z_f$ and the redshift of the GRB host galaxy 
$z_{GRB}$ provides an estimate of the age of the system. 
The best model of the ``expected dust'' scenario points towards an age of 
$\sim 0.6$ Gyr. With the best model of  the ``alternative dust'' scenario, a nearly identical value for 
the age of the GRB host can be derived. \\
At this age, the predicted specific star formation rate is 4.7 Gyr$^-1$, both for the best model of the ``expected dust'' and 
``alternative dust'' scenario. \\
We have also a chance to test whether our assumption for dust evolution are appropriate, in particular the 
hypothesis of having neglected dust accretion. At the age of 0.6 Gyr, with the best model of the ``expected dust'' picture 
we predict a dust-to-gas ratio $D = 1.2 \times 10^{-6}$, which is in very good agreement with the observational upper limit\footnote{To compute the observed dust-to-gas and dust-to-metal ratios, we 
assume for the Small Magellanic cloud a $D_{SMC}=0.1 D_{\odot}=0.0008$ (Issa et al. 1990, Rubio et al. 2004) and $Z_{\odot}=0.012$ 
(Grevesse et al. 2007).}
of Prochaska et al. (2007b) of $8 \times 10^{-6}$.  The predicted dust-to-metals ratio is  $D_Z = 0.03$, again in agreement with the upper limit of 0.12 derived 
observationally by Prochaska et al. (2007b).  Very similar values may be obtained by means of the best model of the 
``alternative dust'' scenario. We conclude that the abundances of this system are not strongly 
influenced by dust depletion. 
Our assumptions concerning the treatment of dust evolution, in particular 
having neglected dust accretion in molecular clouds, seems appropriate for the study of this GRB host.

%%%%%%%%%%%%%%%%%%%%%%%%%%%%%%%%%%%%%%%%%%%%%%%%%%%%%%%%%%%%%%%%%%%%%%%%%%%%%%%%%%%%%%%%%%%%%%%%%%%%%%%%%%%%%%%%%%%
\begin{figure*}
\includegraphics[width=\columnwidth]{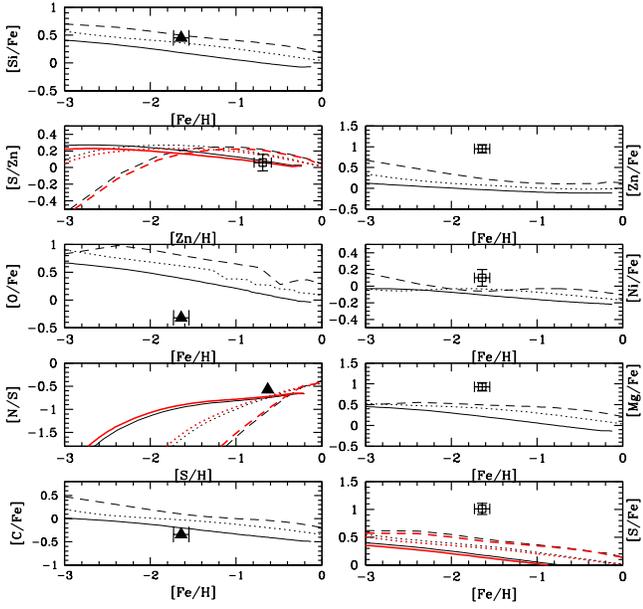}
\caption{ Observed abundance ratios versus metallicity for the host galaxy of GRB 050820
as derived by \citet{PRO07a} (empty squares with error bars and solid triangles). 
The thin solid line, dotted line and dashed line 
are predictions computed by means of a chemical evolution model for 
a dwarf irregular galaxy assuming for the star formation efficiency  
$\nu=0.1 Gyr^{-1}$, $\nu=1 Gyr^{-1}$, and $\nu=10 Gyr^{-1}$, respectively, with ``expected dust'' composition. 
The thick lines are as above, but with ``alternative dust'' composition, i.e. with S depleted. }
\label{fig4}
\end{figure*}
%%%%%%%%%%%%%%%%%%%%%%%%%%%%%%%%%%%%%%%%%%%%%%%%%%%%%%%%%%%%%%%%%%%%%%%%%%%%%%%%%%%%%%%%%%%%%%%%%%%%%%%%%%%%%%%%%%
\begin{figure*}
\includegraphics[width=\columnwidth]{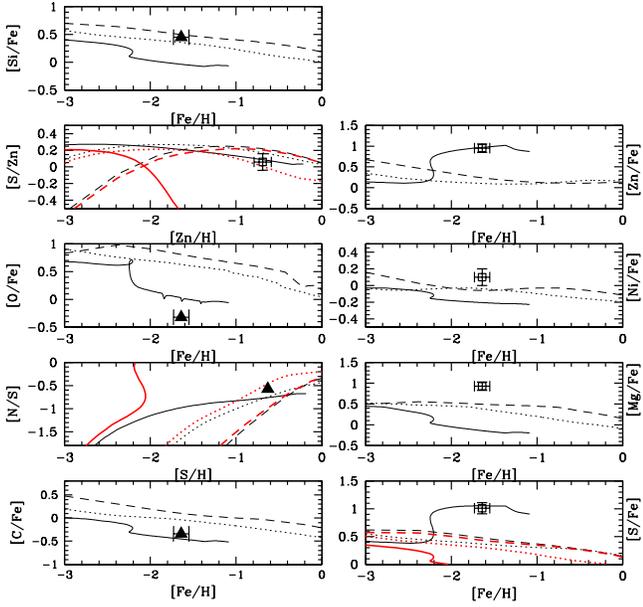}
\caption{Symbols as in Fig 4. Thick and thin lines as in Fig 4, but including dust accretion. }
\label{fig5}
\end{figure*}
%%%%%%%%%%%%%%%%%%%%%%%%%%%%%%%%%%%%%%%%%%%%%%%%%%%%%%%%%%%%%%%%%%%%%%%%%%%%%%%%%%%%%%%%%%%%%%%%%%%%%%%%%%%%%%%%%%
%%%%%%%%%%%%%%%%%%%%%%%%%%%%%%%%%%%%%%%%%%%%%%%%%%%%%%%%%%%%%%%%%%%%%%%%%%%%%%%%%%%%%%%%%%%%%%%%%%%%%%%%%%%%%%%%%%%
\begin{figure*}
\includegraphics[width=\columnwidth]{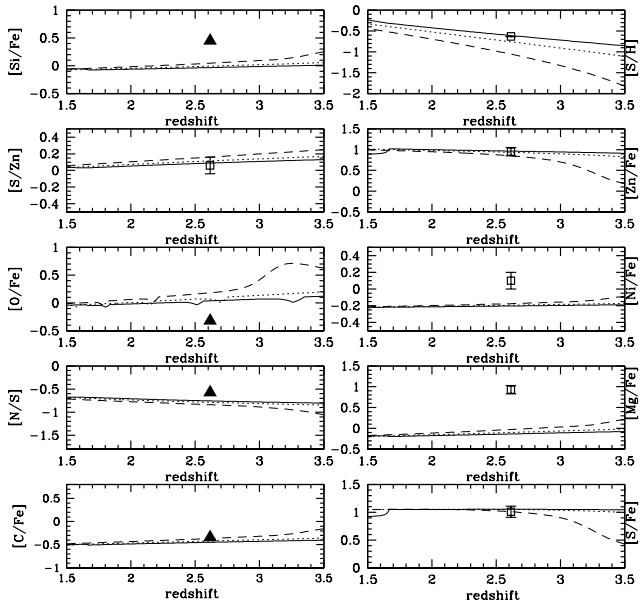}
\caption{Observed abundance ratios versus redshift for the host galaxy of GRB 050820 
as derived by \citet{PRO07a} (empty squares with error bars and solid triangles). 
The dashed line, the dotted line and the solid line 
represent the predictions computed by means of the best model (  $\nu=0.1 Gyr^{-1}$, dust accretion) 
within the ``expected dust'' scenario assuming 
$z_{f}\ge 4$, and $z_{f}=6$, and $z_f=10$, respectively. }
\label{fig6}
\end{figure*}
%%%%%%%%%%%%%%%%%%%%%%%%%%%%%%%%%%%%%%%%%%%%%%%%%%%%%%%%%%%%%%%%%%%%%%%%%%%%%%%%%%%%%%%%%%%%%%%%%%%%%%%%%%%%%%%%%%%

\subsection{GRB 050820}
In fig. 4, we show various observed interstellar 
abundance ratios  vs metallicity for the host galaxy of GRB 050820. 
The observed abundance ratios are compared with predictions computed by means of 
three chemical evolution models with SF efficiencies $\nu=0.1$ Gyr$^{-1}$, $\nu=1$ Gyr$^{-1}$ and 
$\nu=10$ Gyr$^{-1}$. 
In this case, we see how, with a model assuming only dust destruction and 
no dust accretion, the observed abundance ratios do not allow us to determine which model 
is the most accurate in reproducing the data. 
The predictions for  the model with  $\nu=0.1$ Gyr$^{-1}$ allows us to reproduce the [C/Fe], [O/Fe], [S/Zn] and  
are very close to the observed 
[N/S] lower limit. 
However, no single 
model can reproduce the remaining set of abundance ratios shown in Fig. 4. 
In particular, 
the very high [S/Fe], [Mg/Fe] and [Zn/Fe] abundance 
ratios are not reproduced, even by assuming a star formation efficiency $\nu=10$ Gyr$^{-1}$, a value 
generally used to describe elliptical galaxies (Calura, Matteucci \& Vladilo 2003). 
The same conclusion is valid taking into account S depletion (thick lines in Fig. 4). 
This GRB DLA presents peculiar abundance ratios, such as a high value for [Zn/Fe] which, at such 
a [Fe/H], in general is interpreted as the signature of dust depletion (see e.g. Vladilo 2004). 
To investigate the possibility that dust depletion may play a non negligible role in the ISM 
of the GRB 050820 host, we consider a chemical evolution model which takes into account dust accretion. 
To model dust accretion, the prescriptions are the same as described in CPM08. 
The dust accretion rate is $G_{dust,i}/\tau_{accr}$, where $\tau_{accr}$ is the dust accretion timescale
\begin{equation}
\tau_{accr}=\tau_{0,i}/(1 - f_i) 
\label{accr_t}
\end{equation} 
where 
\begin{equation}
f_i=\frac{G_{dust,i}}{G_{i}}
\end{equation} 
For the timescale $\tau_{0,i}$, typical values span from $\sim 5 \times 10^{7}$ yr, of the order of 
the lifetime of a typical molecular cloud, up to $\sim 2 \times 10^{8}$ yr  \citep{DWE98}. 
In this paper, we assume that the timescale $\tau_{0,i}$ is constant for all elements, 
with a value of $5 \times 10^{7}$ yr (CPM08, Inoue 2003). \\
In Fig. 5, we show our results for the predicted abundance ratios of the three models with 
$\nu=0.1$ Gyr$^{-1}$, $\nu=1$ Gyr$^{-1}$ , and $\nu=10$ Gyr$^{-1}$ taking into account also dust accretion. 
By including accretion, a larger number of abundance ratios is  now reproduced by the model with $\nu=0.1$, computed assuming 
no S depletion (thin lines in Fig. 5). In particular, the [S/Fe] and [Zn/Fe] ratios are now reproduced by the model with $\nu=0.1 $ Gyr$^{-1}$. 
The very high [Si/Fe], [Mg/Fe] and [Ni/Fe] are not reproduced by the same model, and they are likely to 
be due to differential depletion effects. 
Differential depletion is the most likely explanation for peculiar [Si/Fe] and [Mg/Fe] ratios observed also 
in some DLAs (Dessauges-Zavadsky et al. 2002, D07). 
In this paper, to limit the  parameter space,  
we have assumed that all the chemical elements are incorporated into dust grains 
in the same proportions. A  detailed modeling  of differential depletion is beyond the aims of the present work.  
Future investigation on this topic, also in connection to dust depletion in DLAs, is required to shed light on the chemical species 
interested by differential depletion. \\
From Fig. 5, we see that, in presence of dust accretion, 
the predicted abundance ratios between a non-refractory element and a refractory one, 
such as [Zn/Fe], are the highest for the model with the lowest 
SF efficiency. This counter-intuitive result is due to the fact that, given the low condensation efficiencies assumed here, 
with high star formation efficiencies ($\nu \ge$ 1 Gyr$^{-1}$), 
dust grains hardly survive in the ISM even if accretion is present. 
This result is due to the fact that a high star formation efficiency implies a high dust destruction rate, 
which  the process of accretion is not able  to compensate. This is consistent with the results of CPM08, 
who found that dust destruction by SNe is more efficient than dust production. 
However, CPM08 assumed the same dust condensation efficiencies as \citet{DWE98}, considerably 
higher than the ones used in the present paper (e.g., for Si and  Fe, CPM08 assumed $\delta_{Si, Fe}=0.8$). 
In this case, also a galaxy with 
a SF efficiency of $\sim 10$ Gyr$^{-1}$, such as an elliptical galaxy of luminous mass $10^{11} M_{\odot}$, can present a very high  dust 
fraction, up to 0.9-1, in the period when star formation is active.\\
It is also interesting to note that, in the ``alternative dust'' scenario, including S depletion and accretion, no model provides a satisfactory fit 
to the observed data. \\
Regarding GRB 050820, we conclude that the model with SF efficiency $\nu=0.1$ Gyr$^{-1}$, ``expected dust'' composition and dust accretion 
is the best in reproducing the observed abundance ratios vs metallicity. We now use this model to 
infer the age of the GRB 050820 host galaxy.\\
In Fig. 6, we show various abundance ratios vs redshift as measured by P07 for GRB 050820, compared to our predictions 
computed by means of the best model and assuming three different redshifts of formation: $z_f=4$, $z_f=6$, and $z_f=10$. 
It is clear that the majority of the abundance ratios, i.e. all the abundance ratios except [N/S], 
[Si/Fe], [Mg/Fe] and [Ni/Fe], all underestimated for the reasons described above, may be reproduced by assuming $z_f \ge 6$. 
In this way, we are able to derive a lower limit of $1.5$ Gyr for the age of the host galaxy of GRB 050820. 
This allows us to derive an upper limit of $2.5$ Gyr$^{-1}$ for the SSFR. Now, we investigate whether our assumptions concerning 
dust provide a realistic description of the dust content of this GRB DLA. 
For the GRB 050820 DLA, Prochaska et al. (2007b) measure a dust-to-gas ratio $0.0008$. Our results imply that, 
at the ages larger or equal to 1.5 Gyr, the predicted dust-to-gas ratio for the best model is $\ge 0.00057$, in very good agreement with the 
observed value. The predicted dust-to-metals ratio is 0.73. 
By assuming for the observed metallicity [Z/H]=[S/H]=-0.67 $\pm 0.1$ and by neglecting te uncertainty in the observed dust-to-gas, 
the measured dust-to-metals ratio is between 0.23 and 0.36. Given the uncertainties of the parameters involved in our computation, in particular the dust condensation 
efficiencies and the accretion timescale, which  may even be larger by a factor of 2 than the value we have adopted, i.e. $\tau_0=0.05$ Gyr, 
we think our models provides a reasonable description of the dust properties of GRB 050820.

%%%%%%%%%%%%%%%%%%%%%%%%%%%%%%%%%%%%%%%%%%%%%%%%%%%%%%%%%%%%%%%%%%%%%%%%%%%%%%%%%%%%%%%%%%%%%%%%%%%%%%%%%%%%%%%%%%%
\begin{figure*}
\includegraphics[width=\columnwidth]{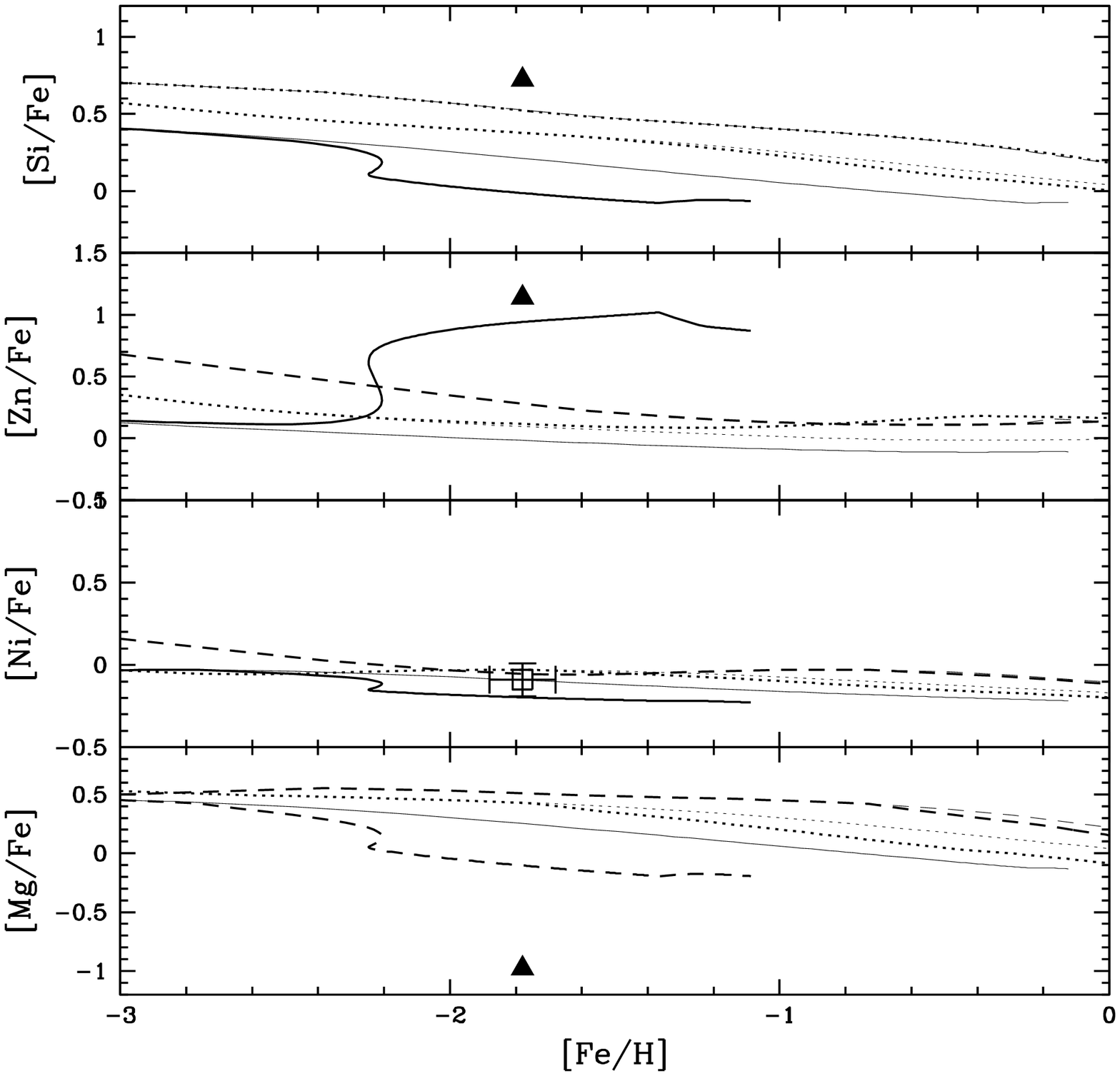}
\caption{ Observed abundance ratios versus metallicity for the host galaxy of GRB 051111
as derived by \citet{PRO07a} (empty squares with error bars and solid triangles). 
The thin solid line, dotted line and dashed line 
are predictions computed by means of a chemical evolution model for 
a dwarf irregular galaxy assuming for the star formation efficiency  
$\nu=0.1 Gyr^{-1}$, $\nu=1 Gyr^{-1}$, and $\nu=10 Gyr^{-1}$, respectively, with ``expected dust'' composition. 
The thick lines are as above, but taking into account dust accretion }
\label{fig7}
\end{figure*}
%%%%%%%%%%%%%%%%%%%%%%%%%%%%%%%%%%%%%%%%%%%%%%%%%%%%%%%%%%%%%%%%%%%%%%%%%%%%%%%%%%%%%%%%%%%%%%%%%%%%%%%%%%%%%%%%%%

\subsection{GRB 051111}
Of the elements studied in this paper, 
for GRB 051111, only 4 measured abundance ratios are available (see Tab. 1). 
However, of these 4, 3 are lower limits and only one is a real measure. 
In fig. 7, we show the observed interstellar 
abundance ratios  vs metallicity for the host galaxy of GRB 051111,  
compared with predictions computed by means of 
three chemical evolution models with SF efficiencies $\nu=0.1$ Gyr$^{-1}$, $\nu=1$ Gyr$^{-1}$ and 
$\nu=10$ Gyr$^{-1}$. The predictions are shown both taking into account dust accretion (thick lines) and 
not (thin lines). The poor number of abundance ratios measured for this GRB DLA does not allow us to constrain 
its star formation history. 
However, the very large [Zn/Fe] is indicative of strong dust depletion. 
Unfortunately, no measures are available for the dust-to-gas and dust-to-metals ratios, 
hence it is not possible to assess accurately the amount of 
dust depletion characterizing this system.
%We await for future observations aimed at 
%extending  the number of abundance measures for this GRB DLA }

%%%%%%%%%%%%%%%%%%%%%%%%%%%%%%%%%%%%%%%%%%%%%%%%%%%%%%%%%%%%%%%%%%%%%%%%%%%%%%%%%%%%%%%%%%%%%%%%%%%%%%%%%%%%%%%%%%%
\begin{figure*}
\includegraphics[width=\columnwidth]{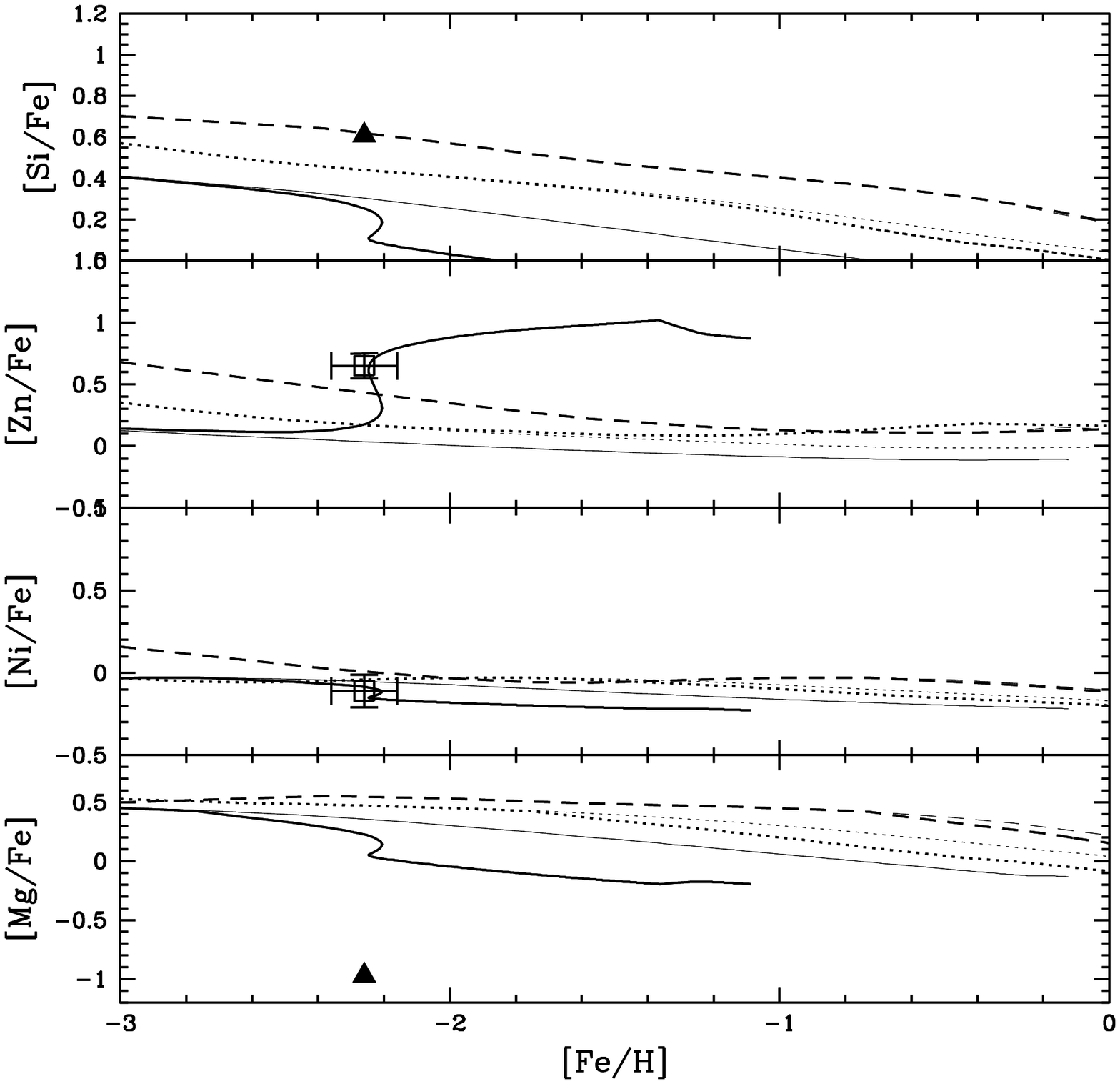}
\caption{ Observed abundance ratios versus metallicity for the host galaxy of GRB 060418
as derived by \citet{PRO07a} (empty squares with error bars and solid triangles). 
The thin solid line, dotted line and dashed line 
are predictions computed by means of a chemical evolution model for 
a dwarf irregular galaxy assuming for the star formation efficiency  
$\nu=0.1 Gyr^{-1}$, $\nu=1 Gyr^{-1}$, and $\nu=10 Gyr^{-1}$, respectively, with ``expected dust'' composition. 
The thick lines are as above, but taking into account dust accretion }
\label{fig8}
\end{figure*}
%%%%%%%%%%%%%%%%%%%%%%%%%%%%%%%%%%%%%%%%%%%%%%%%%%%%%%%%%%%%%%%%%%%%%%%%%%%%%%%%%%%%%%%%%%%%%%%%%%%%%%%%%%%%%%%%%%
\begin{figure*}
\includegraphics[width=\columnwidth]{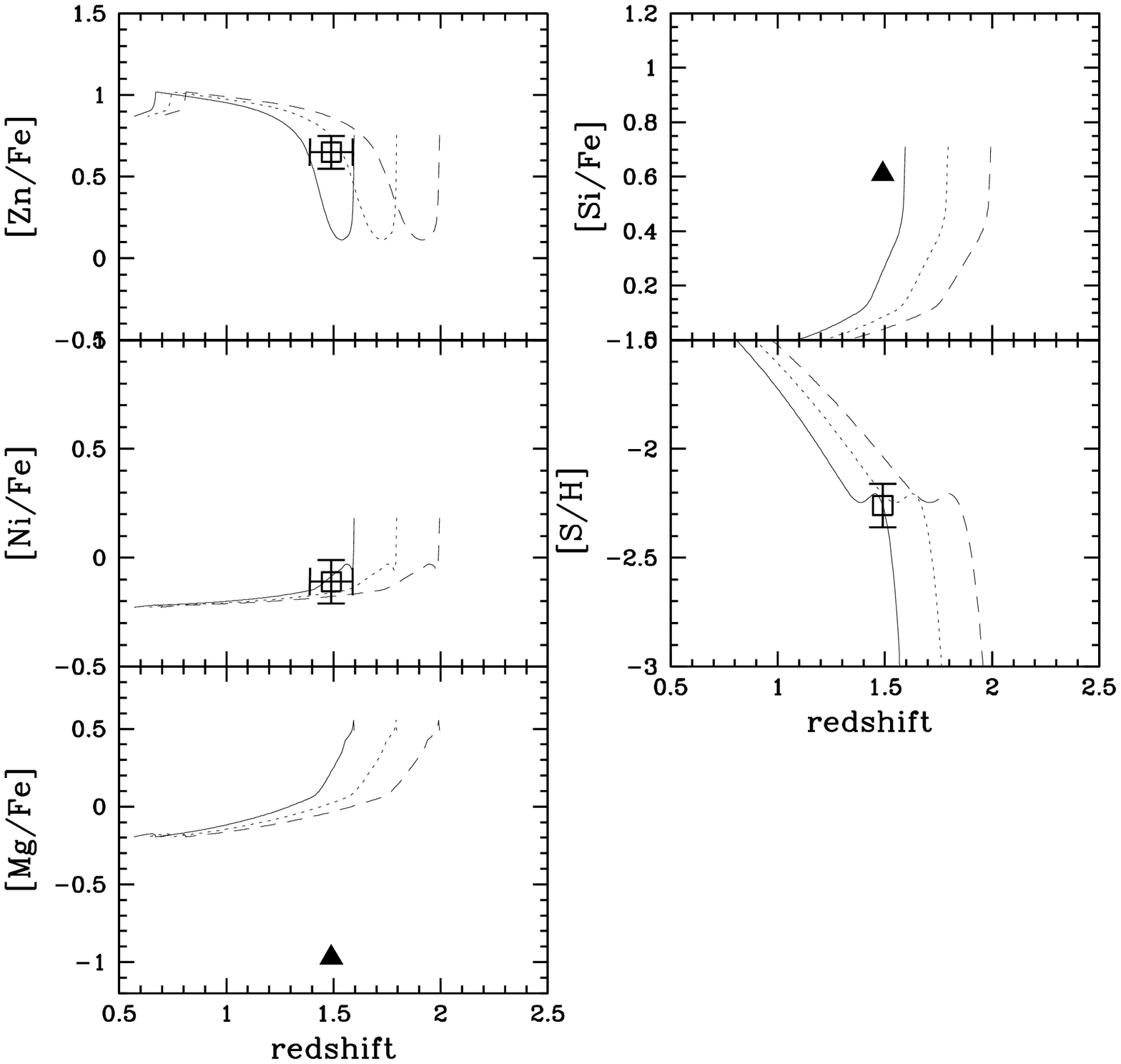}
\caption{Observed abundance ratios versus redshift for the host galaxy of GRB 060418
as derived by \citet{PRO07a} (empty squares with error bars and solid triangles). 
The solid line, the dotted line and the dashed line 
represent the predictions computed by means of the best model (  $\nu=0.1 Gyr^{-1}$, dust accretion) 
within the ``expected dust'' scenario assuming 
$z_{f}\ge 1.6$, and $z_{f}=1.8$, and $z_f=2$, respectively. }
\label{fig9}
\end{figure*}
%%%%%%%%%%%%%%%%%%%%%%%%%%%%%%%%%%%%%%%%%%%%%%%%%%%%%%%%%%%%%%%%%%%%%%%%%%%%%%%%%%%%%%%%%%%%%%%%%%%%%%%%%%%%%%%%%%

\subsection{GRB 060418}
Also For GRB 060418, only 4 measured abundance ratios are available to be compared with our predictions.  
In fig. 8, we show the observed interstellar 
abundance ratios  vs metallicity for the host galaxy of GRB 060418,  
compared with predictions computed by means of 
three chemical evolution models with SF efficiencies $\nu=0.1$ Gyr$^{-1}$, $\nu=1$ Gyr$^{-1}$ and 
$\nu=10$ Gyr$^{-1}$, taking into account dust accretion (thick lines) and 
not (thin lines). 
In this case, we have two real measures and two lower limits.
Also this system presents a supersolar [Zn/Fe] ratio, which is likely indicating the 
presence of dust depletion. The two actual measures allow us to derive some constraints 
on the star formation history and on the age of the host galaxy of GRB 060418. 
From Fig. 8, we see that the model which best reproduces the observed abundance ratios 
is the one characterized by $\nu=0.1$ Gyr$^{-1}$ and including dust accretion. 
With this model, it is possible to reproduce two measured 
abundance ratios, the lower limit for [Mg/Fe] is compatible with our predictions, whereas 
the [Si/Fe] is not, probably owing to the effects of differential depletion, as explained above. 
We use this model to derive a value for the age and for the SSFR of the host galaxy of this GRB. \\
In Fig. 9, we show the measured abundance ratios vs redshift, compared to our predictions
computed with the best model, assuming three different values for the redshift of 
formation $z_f$: $z_f=1.6$,  $z_f=1.8$ and $z_f=2.$. The observed abundance ratios are reproduced 
by assuming $z_f=1.8$. 
The redshift of the GRB is $z_{GRB}=$1.5, which, with the cosmology adopted in this paper,  
implies an age of 0.63 Gyr and a SSFR of 2.8 Gyr$^{-1}$. 
At  this age, we predict a dust-to-gas ratio of 0.0005 and a dust-to-metals ratio of 0.71. 
Unfortunately, 
no observational value for the dust-to-gas ratio is available for this system. \\

A summary of the predicted physical properties of the GRB host galaxies studied in this paper is presented in Tab. 2.

\subsection{Specific star formation rates and implications for cosmic chemical evolution studies}
\begin{figure*}[t]
\includegraphics[height=20pc,width=20pc]{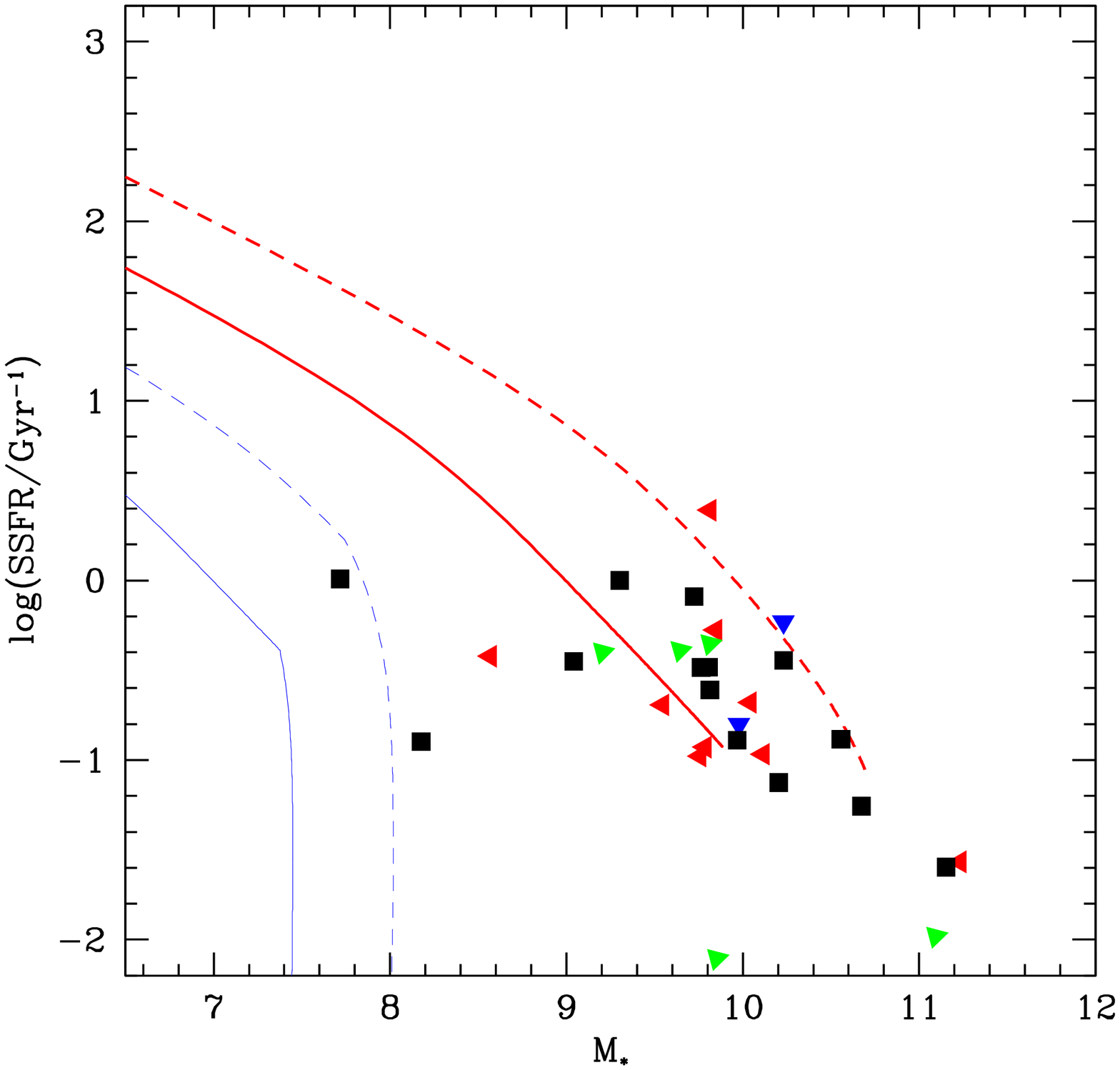}
\includegraphics[height=20pc,width=20pc]{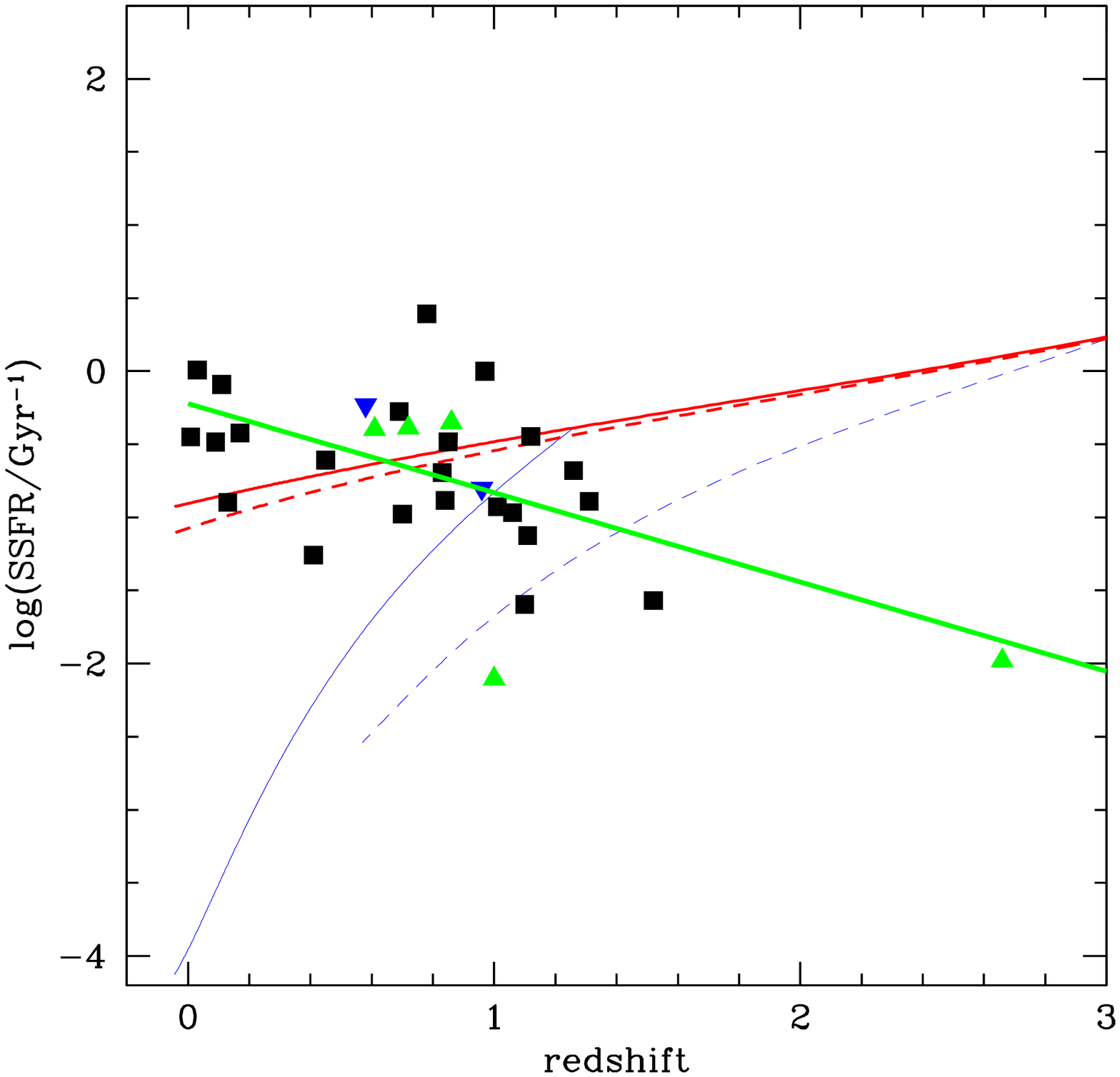}
\caption{\emph{(a)}:  SSFR as a function of the stellar mass 
for two models 
with SF efficiency $\nu=0.01$ Gyr$^{-1}$ (solid lines) and $\nu=0.1$ Gyr$^{-1}$ (dashed lines), and assuming 
$M_{tot}=10^9 M_{\odot}$ (thin lines) and $M_{tot}=10^{11} M_{\odot}$ (thick lines).
The solid squares and the triangles are the observational determinations by  \citet{CAS08}.  The triangles mark lower 
or upper limits for the derived stellar mass, SSFR or both.
\emph{(b)}: Redshift evolution of the SSFR for our models, 
and for the GRB host galaxies as observed  by \citet{CAS08}. All the curves, squares and triangles are as in panel $a$. 
The thick dot-dashed line represents the best fit to the observed values, 
computed by means of a straight line $y=ax+b$. All the other curves, squares and triangles are as in panel $a$. }
\label{fig10}
\end{figure*}

The specific star formation rate (SSFR) is defined as the 
ratio between the SFR and the stellar mass.
Its determination does not require any a-priori assumption on unknown quantities 
involved in our study, 
such as the mass or the physical dimension of the system, which on the other hand have to be assumed 
in order to provide an estimate of the SFR or  the SFR surface density.\\ 
The SSFR provides an indication of the intensity of the 
star formation of the host galaxy, whereas the inverse of the 
SSFR represents the star formation timescale, hence a high value for the SSFR 
is associated to a young galaxy \citep{CAS06} . 
It is interesting to compare the SSFR values predicted by means of our models  with the SSFR observed for a 
set of GRB host galaxies at redshifts $0<z<2.7 $ by \citet{CAS08}. 
\citet{CAS08} used the Spitzer images of 30  GRB hosts and estimated their stellar masses 
and their unobscured star 
formation rates by means of their K-band fluxes and rest-frame UV spectra, respectively. 
By means of these data, it has been possible to determine the SSFRs for the host galaxies of their sample.  
%In Fig--, upper panel, we show the time evolution of the SSFRs for 
%the minimal and maximal models computed within the ``expected dust'' scenario.  
%Note that the the predictions for the two models coincide, hence, in this time interval, 
%the SSFR is not dependent on the SF efficiency $\nu$. This is due to the fact that,
% to a  first approximantion, both the SFR and the stellar mass are 
%proportional to the SF efficiency $\nu$. 
%In general, the values predicted by means of our models are larger 
%than the range of values found by \citet{CAS08}, which, as discussed below, in most of the cases should represent 
%lower limits. \\
In the left panel of Fig.~\ref{fig10}, we show the SSFR as a function of the stellar mass for two models 
with SF efficiency $\nu=0.01$ Gyr$^{-1}$ (solid lines) and $\nu=0.1$ Gyr$^{-1}$ (dashed lines). 
These models have turned out to be the best in describing the chemical abundance patterns of the GRB DLAs studied in this paper. 
The models shown in Fig. 10 assume various values for the mass normalization $M_{tot}$:
$M_{tot}=10^9 M_{\odot}$ (thin lines) and $M_{tot}=10^{11} M_{\odot}$ (thick lines). This stellar mass 
range brackets the values observed in the bulk of the data by \citet{CAS08}. The models are compared to the data by  \citet{CAS08}. \\
%Note that each model plotted in Fig.~\ref{fig4} is characterized by the same chemical evolution and, 
%at every timestep until the age of the universe at $z=z_{GRB}$, 
%by  the same abundance ratios. 
%The same is true also for the three curves for the minimal models 
%plotted in Fig. ~\ref{fig4}. 
Until the onset of the galactic wind, 
each pair of models with the same SF efficiency plotted in Fig.~\ref{fig10} is characterized by the same chemical evolution and  
by  the same abundance ratios. 
This is due to the fact that in chemical evolution models, 
to a first approximation,  
the abundance ratios do not depend on the total mass of the system, which  
plays a key role only in the presence of outflows. \\
%All the models used here experience winds at times much larger than 1.5 Gyr, i.e. the age of the Universe 
%at redshift $z=z_{GRB}$. \\
The predictions shown in Fig. ~\ref{fig10} have been computed until the present time, corresponding 
to $\sim 13.5$ Gyr for the cosmology adopted in the present paper. This is done to be consistent with the fact 
that the redshift range spanned by the data   by  \citet{CAS08} is large, encompassing $\sim 82 \%$ of the cosmic time. 
Several observational values plotted in Fig. ~\ref{fig10}  represent upper limits for the stellar mass and lower limits for the SSFR. 
In fact, \citet{CAS08} showed that dust extinction must be present in at least $25\%$ of the GRB hosts, 
causing the SFR determinations derived from UV/optical observations to be regarded as lower limits. 
The SSFR values by \citet{CAS08} should be regarded as lower limits to the actual values also owing to 
a ``dilution'' effect present in mid-infrared photometry studies. 
This effect is related 
to the fact that some hosts of the sample are not spatially resolved, hence, the SSFRs are computed by dividing the 
UV SFRs by the total stellar masses, instead of dividing the SFRs by the stellar masses of the star forming regions, 
which are likely to encompass a small part of the host galaxy. 
For these reasons,  if the corrections for dust extinction and dilution  
were performed  and if the stellar masses were determined with higher precision, 
at least half of the data of \citet{CAS08} should move towards our predictions. \\
In Fig. ~\ref{fig10}, right panel, we show the redshift evolution of the SSFR for our models, 
and for the GRB host galaxies as observed  by \citet{CAS08}. 
In this case, in all our models, we have assumed that star formation begins at the same redshift $z=5$.   
Some of the predictions for the various models coincide across a large redshift interval.  
This is due to the fact that the SSFR is not dependent on the SF efficiency $\nu$ since, 
 to a  first approximation, both the SFR and the stellar mass are 
proportional to the SF efficiency $\nu$. 
In some cases, the predictions diverge owing to the onset of a galactic wind. 
This event concerns the models with the lowest mass normalizations, represented by the two thin lines 
in the right panel of Fig.~\ref{fig10},   
characterized by the shallowest gravitational potentials. As soon as the galactic wind develops, 
a considerable amount of  gas becomes unavailable for star formation. 
The time the wind develops depends on the mass of the system and on the star formation efficiency $\nu$. 
In general, the higher is the adopted $\nu$  and the lower is the total mass, the earlier the wind develops. 
For this reason, after the onset of the wind, the SSFR does depend on the adopted  star formation efficiency $\nu$.\\
According to our predictions, the SSFRs decrease with decreasing redshift. 
Our result in not surprising, since any galactic evolution model 
predicts an increase of the stellar mass and a decrease of the SFR  with decreasing redshift. 
This is in agreement also with the observed cosmological evolution of the SFR, in particular with  
the observed decline of the cosmic star formation history with redshift at $z<2$
(Hopkins\& Beacom 2006), likely due to progressive gas consumption in late-type spiral discs \citep{CAL03a, CAL06}. 
Furthermore, \citet{PAP06} showed that the integrated specific SFR, defined as the ratio between 
the comoving SFR density and the comoving stellar mass density, is a decreasing function of reshift. 
These results are apparently in contrast with the observed redshift evolution of the SSFR in 
GRB host galaxies. In fact, the  data by \citet{CAS08} suggest an increase of the SSFRs with decreasing redshift 
(see the solid straight line plotted in ~\ref{fig10}), at variance 
with any results concerning the redshift evolution of star forming galaxies at redshifts $0 \le z \le 3$. 
%This fact is likely due to selection effects present in the sample by \citet{CAS08}, but certainly does  
%cast serious doubts on the fact that GRB host galaxies may represent unbiased tracers of star formation at any redshift \citep{LEF06}. 
As stated also by \citet{CAS08}, the GRB host galaxies with the most vigorous star formation 
are likely to be heavily dust obscured, preventing the localization of their afterglows. 
This is one possible reason why, in the right panel  of Fig.~\ref{fig10}, 
no objects lay in the right upper corner of the plot. 
In the future, it will be important to extend the samples of GRB hosts with known star formation and stellar mass measures, 
in order to understand the reason for this apparent discrepancy and 
understand whether this trend is  peculiar to GRB hosts or due to some selection effect. 
Surveys are ongoing 
to comprise GRB host galaxy samples based on localizations
using the X-ray afterglow which is less susceptible to dust (Fynbo et al. 2007).
%Our study shows that, owing to their detection technique 
%and their samples being biased against dust-extincted systems, it may be 
%difficult to trace highly obscured star formation with
%optically selected GRB samples.  
%Current GRB host galaxy samples 
%have required the detection of an optical afterglow to localize
%the event.  This criterion
%biases the sample against events that occur in extremely dust
%regions of SF galaxies.  In turn, this may bias the sample against
%the highest SF systems.  Surveys are ongoing 
%to comprise GRB host galaxy samples based on localizations
%using the X-ray afterglow which is less susceptible to dust.
%GRB host galaxies are unlikely to map the 
%bulk of the star formation at any redshift and are not representative of all star-forming galaxies at high redshift. \\

%%%%%%%%%%%%%%%%%%%%%%%%%%%%%%%%%%%%%%%%%%%%%%%%%%%%%%%%%%%%%%%%%%%%%%%%%%%%%%%%
\begin{table*}
\caption[]{Summary of the physical properties of the 4 GRB host galaxies studied in the present paper}
\begin{tabular}{lcccc}
\\[-2.0ex] 
\hline
\\[-2.5ex]
%             &                     &                    &           &           &  GRB 050730  &      \\ 
%multicolumn{1}{c}{}&\multicolumn{1}{c}{}&\multicolumn{1}{c}{}&\multicolumn{1}{c}{}&\multicolumn{1}{c}{}&\multicolumn{1}{c}{}&\multicolumn{1}{c}{GRB 050730}&\multicolumn{1}{c}{}\\
\hline
%\multicolumn{1}{c}{}&\multicolumn{1}{c}{Observed}&\multicolumn{1}{c}{Predicted}&\multicolumn{1}{c}{Predicted}\\
%\multicolumn{1}{c}{}&\multicolumn{1}{c}{}&\multicolumn{1}{c}{(Minimal M.)}&\multicolumn{1}{c}{(Maximal M.)}\\
%\multicolumn{1}{c}{}&\multicolumn{1}{c}{(MW)}&\multicolumn{1}{c}{(SMC/LMC)}&\multicolumn{1}{c}{}&\multicolumn{1}{c}{best min.}&\multicolumn{1}{c}{best max.}&\multicolumn{1}{c}{[X/H]}\\
%\multicolumn{1}{c}{}&\multicolumn{1}{c}{}&\multicolumn{1}{c}{}&\multicolumn{1}{c}{}&\multicolumn{1}{c}{}&\multicolumn{1}{c}{}&\multicolumn{1}{c}{GRB 050730}&\multicolumn{1}{c}{GRB 050730}\\
%\multicolumn{1}{c}{}&\multicolumn{1}{c}{observed}&\multicolumn{1}{c}{pred.}&\multicolumn{1}{c}{pred.}\\
%\multicolumn{1}{c}{}&\multicolumn{1}{c}{(P07)}&\multicolumn{1}{c}{(no dust)}&\multicolumn{1}{c}{(dust)}\\
\multicolumn{1}{c}{GRB}&\multicolumn{1}{c}{Age ($Gyr$)} &\multicolumn{1}{c}{SSFR ($Gyr^{-1}$)} & \multicolumn{1}{c}{Dust-to-gas} & \multicolumn{1}{c}{Dust-to-metals}\\
050730    &  0.61      & 4.7                 &  $1.2 \times 10^{-6}$     &  0.03 \\
050820    &  $\ge$1.53 & $\le 2.5$ Gyr$^{-1}$ &   $\ge 0.00057$          &  0.73 \\
051111    &   -        & -  & -   & -   \\
060418    &  0.66      & -  & 0.0005   &  0.71  \\
%\\[-1.0ex]2.8 Gyr$^{-1}$. 
\hline
%QB0841+129    &     1.5    \\               % bab6  
%PKS1157+014   &     0.57    \\              % bap1
%QB1210+175    &     0.48     \\             % bac1
%QB2230+02     &     0.24      \\            % bar1                     
%QB2348-1444   &      7.        \\              % bbt               
%\hline
%\multicolumn{1}{c}{GRB }&\multicolumn{1}{c}{SSFR}\\
%\hline
%\\[-1.0ex]
%GRB 050730  &      1.3     \\
%            &    (minimal E.D.) \\
%            &     5.44    \\
%            &    (maximal   E.D.)   \\
%            &     21.8   \\
%            &    (A. D.)   \\
%[N/S]               &   -0.93   &  -0.86   & -0.95  \\
%$[$S/Fe$]$          &    0.23   &  0.29    & 0.32   \\ 
%$[$Ni/Fe$]$         &   -0.16   &  -0.12    & -0.07   \\
%S            &   0$^{2}$           &  ...               &  &0         &  0     &  -2.2  \\ 
%Fe           &   0.65-1$^{1,2}$    &  0.3-1$^{3,4,5}$    &  &0.49      &  0.32   &  -2.43\\ 
%Ni           &   0.79-1$^{1}$      &  0.3-1$^{3,4,5}$    &   &0.49      & 0.34   &  -2.59\\ 
\hline
\hline
\end{tabular}
\label{tab1}
\end{table*}
%%%%%%%%%%%%%%%%%%%%%%%%%%%%%%%%%%%%%%%%%%%%%%%%%%%%%%%%%%%%%%%%%%%%%%%%%%%%%%%%%%%%%%%%%%%

%%%%%%%%%%%%%%%%%%%%%%%%%%%%%%%%%%%%%%%%%%%%%%%%%%%%%%%%%%%%%%%%%%%%%%%%%%%%%%%%
\begin{table*}
\caption[]{Specific Star Formation Rates predicted for GRB hosts and QSO DLAs}
\begin{tabular}{lc}
\\[-2.0ex] 
\hline
\\[-2.5ex]
%             &                     &                    &           &           &  GRB 050730  &      \\ 
%multicolumn{1}{c}{}&\multicolumn{1}{c}{}&\multicolumn{1}{c}{}&\multicolumn{1}{c}{}&\multicolumn{1}{c}{}&\multicolumn{1}{c}{}&\multicolumn{1}{c}{GRB 050730}&\multicolumn{1}{c}{}\\
\hline
%\multicolumn{1}{c}{}&\multicolumn{1}{c}{Observed}&\multicolumn{1}{c}{Predicted}&\multicolumn{1}{c}{Predicted}\\
%\multicolumn{1}{c}{}&\multicolumn{1}{c}{}&\multicolumn{1}{c}{(Minimal M.)}&\multicolumn{1}{c}{(Maximal M.)}\\
%\multicolumn{1}{c}{}&\multicolumn{1}{c}{(MW)}&\multicolumn{1}{c}{(SMC/LMC)}&\multicolumn{1}{c}{}&\multicolumn{1}{c}{best min.}&\multicolumn{1}{c}{best max.}&\multicolumn{1}{c}{[X/H]}\\
%\multicolumn{1}{c}{}&\multicolumn{1}{c}{}&\multicolumn{1}{c}{}&\multicolumn{1}{c}{}&\multicolumn{1}{c}{}&\multicolumn{1}{c}{}&\multicolumn{1}{c}{GRB 050730}&\multicolumn{1}{c}{GRB 050730}\\
%\multicolumn{1}{c}{}&\multicolumn{1}{c}{observed}&\multicolumn{1}{c}{pred.}&\multicolumn{1}{c}{pred.}\\
%\multicolumn{1}{c}{}&\multicolumn{1}{c}{(P07)}&\multicolumn{1}{c}{(no dust)}&\multicolumn{1}{c}{(dust)}\\
\multicolumn{1}{c}{DLA}&\multicolumn{1}{c}{SSFR ($Gyr^{-1}$)}\\
\\[-1.0ex]
\hline
QB0841+129    &     1.5    \\               % bab6  
PKS1157+014   &     0.57    \\              % bap1
QB1210+175    &     0.48     \\             % bac1
QB2230+02     &     0.24      \\            % bar1                     
QB2348-1444   &      7.        \\              % bbt               
\hline
\multicolumn{1}{c}{GRB }&\multicolumn{1}{c}{SSFR ($Gyr^{-1}$)}\\
\hline
%\\[-1.0ex]
GRB 050730  &      4.7     \\
GRB 050820  &     $\le$ 2.5   \\
GRB 050820  &     $\le$ 2.8   \\
%[N/S]               &   -0.93   &  -0.86   & -0.95  \\
%$[$S/Fe$]$          &    0.23   &  0.29    & 0.32   \\ 
%$[$Ni/Fe$]$         &   -0.16   &  -0.12    & -0.07   \\
%S            &   0$^{2}$           &  ...               &  &0         &  0     &  -2.2  \\ 
%Fe           &   0.65-1$^{1,2}$    &  0.3-1$^{3,4,5}$    &  &0.49      &  0.32   &  -2.43\\ 
%Ni           &   0.79-1$^{1}$      &  0.3-1$^{3,4,5}$    &   &0.49      & 0.34   &  -2.59\\ 
\hline
\hline
\end{tabular}
\label{tab1}
\end{table*}
%%%%%%%%%%%%%%%%%%%%%%%%%%%%%%%%%%%%%%%%%%%%%%%%%%%%%%%%%%%%%%%%%%%%%%%%%%%%%%%%%%%%%%%%%%%

%%%%%%%%%%%%%%%%%%%%%%%%%%%%%%%%%%%%%%%%%%%%%%%%%%%%%%%%%%%%%%%%%%%%%%%%%%%%%%%%%%%%%%%%%%%%%%%%%%%%%%%%%%%%%%%%%%
%%%%%%%%%%%%%%%%%%%%%%%%%%%%%%%%%%%%%%%%%%%%%%%%%%%%%%%%%%%%%%%%%%%%%%%%%%%%%%%%%%%%%%%%%%%%%%%%%%%%%%%%%%%%%%%%%%%
\begin{figure*}
\includegraphics[width=\columnwidth]{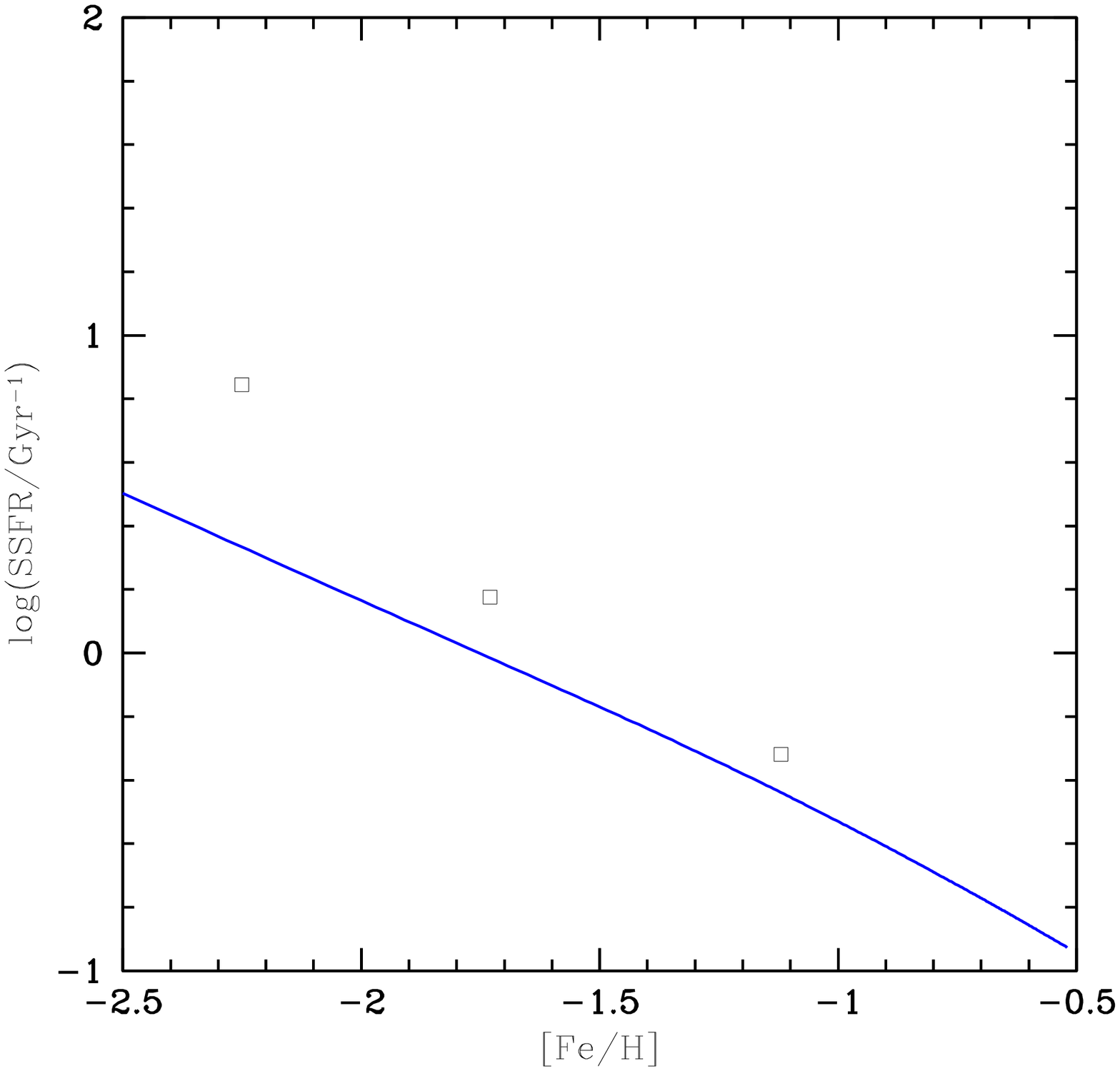}
\caption{Predicted evolution of the SSFR vs [Fe/H] for the best model for GRB 050730 host galaxy. 
The open squares are the values 
determined for the three QSO DLAs of  the sample by \citet{DES07} presenting the lowest dust depletions.}
\label{fig11}
\end{figure*}
%%%%%%%%%%%%%%%%%%%%%%%%%%%%%%%%%%%%%%%%%%%%%%%%%%%%%%%%%%%%%%%%%%%%%%%%%%%%%%%%%%%%%%%%%%%%%%%%%%%%%%%%%%%%%%%%%%

\subsection{A Comparison with SSFRs of Quasar DLAs}
In this section, we compare the SSFR of our models for GRB hosts   
with the values determined by means of  the chemical evolution models for  QSO DLAs presented by \citet{DES07}. 
We focus on the set of dwarf irregular galaxy models characterized by a continuous star formation 
history, which best reproduce the abundances of the set of DLAs studied by  \citet{DES07}. 
%We use the set of chemical evolution models described in Table 2 of Dessauges-Zavadsky et al. (2007) for 
%dwarf irregular with continuous star formation to 
%The SSFRs for 5 DLAs of the sample of  Dessauges-Zavadsky et al. (2007) are calculated 
%for the dwarf irregular models with continuous star formation  which could reproduce at best 
%the abundance patterns observed in --- (see Table --- of Dessauges-Zavadsky et al. 2007).
By means of these models, the SSFRs are calculated at the time corresponding to the age predicted 
for  the DLAs. In this way, for the 5 DLAs towards QSOs Q B0841+129, PKS 1157+014, Q B1210+175, Q B2230+02, 
Q B2348-1444, we find SSFR between 0.24 Gyr$^{-1}$ and 7 Gyr$^{-1}$. 
In Table 3, we show the SSFR values computed for the 5 DLAs of \citet{DES07}, compared to the values 
obtained for the 
hosts of GRB 050730. 
The SSFR computed for the DLAs studied by  \citet{DES07} are in substantial agreement with independent 
estimates for a larger sample of DLAs analysed by \citet{HEN07}.   \\
Although the technique used by \citet{DES07} 
to correct the observations for dust depletion is different than the ones 
used here, it is interesting to note that the SF efficiencies of the 
models used to describe the abundances of QSO DLAs range from $0.05 Gyr^{-1}$ to $0.1 Gyr^{-1}$ 
(see Table 7 of Dessauges-Zavadsky et al. 2007), i.e. basically similar to the values used here.   
%However, the difference in the undepleted chemical evolution pattern resulting from models with 
%$\nu = 0.01 Gyr^{-1}$ and $\nu = 0.1 Gyr^{-1}$ are in general very little, varying by less than 0.1 dex. 
A quantitative analysis of the differences in  the dust depletion corrections computed here and 
the ones used by \citet{DES07} is beyond the scope of the present paper, but it is certainly an interesting 
subject for future studies.\\

It is  may be interesting to compare the SSFR values for the QSO DLAs of D07 with the values found in this paper 
for GRB DLAs. 
The abundances of the DLAs studied by D07 required various 
depletion corrections. 
To correct for dust depletions, D07 used the technique described by Vladilo (2004). The implied depletion 
corrections (considering the [Fe/H] abundance) 
were 0.14 dex for QB0841+129 at z=2.375, 0.28 dex for 
PKS1157+014, 0.14 dex for QB1210+175, 0.33 dex for  QB2230+02  and 0.12 dex for QB2348-1444. \\
The approach to deal with dust depletion used here is different than the one by 
D07. However, the two approaches are consistent, since both predict high [Zn/Fe] ratios for extremely dusty systems, and 
solar or nearly solar [Zn/Fe] for dust-free systems. Moreover, possible depletion effects 
involving S do not influence our results (see Sect.~\ref{050730}). \\
It is probably wise and cautious to compare the SSFRs obtained for the three most dust poor DLAs of D07, 
with the star formation history derived here for GRB 050730, for which both the observations and our predictions 
indicate a very low dust content. 
To perform the comparison, we choose the SSFRs derived for QB0841+129 at z=2.375, QB1210+175 and 
QB2348-1444. The values found for these 3  QSO DLAs of \citet{DES07} are plotted in Fig.~\ref{fig11}. 
As can be seen from Fig.~\ref{fig11}, 
the SSFR of these 3 QSO DLAs are higher than  the  SSFRs of the best models of the ``expected dust'' scenario 
(within the ``alternative dust'' scenario, a very similar curve is obtained). 
%but much lower than the SSFR values of the best model of the ``alternative dust'' scenario. 
%The constraints provided by the models of the ``expected dust'' scenario indicate a noticeable 
%similarity between the SFHs of the QSO DLAs and of the GRB host galaxy studied in this paper. 
%On the other hand, the  ``alternative dust'' scenario points towards much higher star formation efficiencies 
%for the host galaxy of GRB with respect to the values found for QSO DLAs.  \\
The majority of QSO-DLA sightlines
do not penetrate highly SF galaxies, being the most metal rich, 
most vigorously star-forming systems very rare in QSO DLA samples, possibly owing 
to dust obscuration of the background quasars \citep{HOP06, VLA08}. 
Current expectation is that QSO-DLAs, which are drawn according
to HI cross-section, most frequently correspond
to sub-$L_{*}$ galaxies, as evidenced by their significantly sub-solar metallicities 
(e.g. Fynbo et al. 2008).   
At high z, the luminosity function is sufficiently
steep that sub-$L_{*}$ galaxies dominate the integrated SFR \citep{FYN08}. 
For this reason, DLA samples should be able to trace the bulk of star formation. \\
\citet{PRO07a} discuss how GRB DLAs preferentially probe gas associated to the innermost 
regions of galaxies. This gas is generally denser and more metal rich 
than the gas probed by QSO DLAs which, on the other hand, are likley to arise 
in the outer galactic regions. 
If GRB DLAs probe regions of the Universe different than the ones of QSO DLAs,   
the GRB DLA and QSO DLA samples may provide complementary information on cosmic chemical evolution. 

\section{Conclusions}
%The afterglow of the long GRB 050730 detected by the Swift satellite has allowed us 
%to derive accurate column density and elemental abundance 
%measurements for various chemical species present 
%in the ISM of its galaxy host. 
In this paper, for the first time we have used the elemental abundances observed in the host galaxy of 
4 long GRBs  to 
constrain  
their star formation history  and its age.  
Our method is based on the  simultaneous study of  the abundance
ratios between various elements synthesized by stars on different timescales, 
by comparing the measured abundance ratios to the predictions of a detailed chemical evolution model.
% we were able to 
%gain crucial information on the nature of the GRB host galaxy. 
From the simultaneous study of the abundance ratios between various elements vs metallicity (traced by [Fe/H], [S/H] or [Zn/H]), 
we have been able to constrain 
the star formation history of some of the host galaxies. 
By studying the abundance ratios vs redshift, we have been able to constrain 
the redshift of formation and the age of  the GRB host galaxies. We tested models both with and without dust accretion. 
Some peculiar abundance ratios, such as very high [Zn/Fe] ratios, seem to indicate severe dust depletion and, to be explained,  
require the inclusion of dust accretion in the models. 
Our main results can be summarized as follows: \\
1) Our results point towards star formation efficiencies between $0.01Gyr^{-1}$ and $0.1Gyr^{-1}$, 
ages for the system between 0.6 and $\ge$1.5  Gyr. The observed dust content of some GRB DLAs is also satisfactorily accounted for. \\
2) The predicted specific star formation rate values  are between 2.5 $Gyr^{-1}$ 
and 4.7 $Gyr^{-1}$. 
These values are compatible 
with observational estimates from a very recent compilation by \citet{CAS08}. 
Our models predict a decrease of the SSFR with redshift, consistent with the observed decrease of the comoving 
cosmic SFR density between $z \sim2 $ and $z=0$, in agreement with numerous observations of star forming galaxies at low and high redshift. 
On the other hand, apparently the detected GRB hosts follow an opposite trend in the SSFR vs redshift plot, 
with a slight increase of the SSFR with decreasing redshift. 
In the future, it will be important to asses whether this apparent trend is real or if it is due to some selection effect 
affecting the observation of GRB host galaxies. \\
3) The study of the abundance pattern of the host galaxy of GRB 050730 indicates that 
the predicted specific star formation rate is compatible 
with the values found for the 
the set of DLAs studied by \citet{DES07}. 
This implies that, 
the host galaxy of GRB 050730 may have followed a chemical evolution path similar to the ones typical of QSO DLAs.\\  
%from a chemical evolution point of view, 
%this GRB host and its star formation history may have been  not different from  the ones typical of QSO DLAs. 
Our study favors the hypothesis that long duration GRBs occur preferentially in 
low metallicity, star forming galaxies, characterized by low star formation efficiencies ($\nu \le 0.1$ Gyr$^{-1}$). \\ 
%This is supported by the fact that 
%most of the metallicity measurements of GRB host galaxies showed metallicities rarely higher than 0.5 solar --SAVAGLIO 08--. %\citep{BER07}.  
%As for DLAs, the metallicity of the GRB host galaxies is not changing substantially as a function of redshift 
%(Fynbo et al. 2006, and Prochaska et al. 2007)---. 
%\citep{PRI07}. 
The fact that GRBs occur preferentially in low metallicity systems is indicated  also by the luminosity-metallicity relation 
\citep{WOL07}
and by some progenitor models for GRBs, where low metallicity hampers large loss of angular momentum  and mass  
and the progenitor can retain a rapid rotating core, as required in particular in the collapsar model 
for long GRBs \citep{MAC99}. \\
In the future, in order to 
have a better understanding of the chemical evolution patterns observed in high-redshift galaxies, 
it will be important to refine the study of dust depletion in local and distant galaxies, 
in order to clarify whether elements such as S have a refractory nature or not. \\
Furthermore, it will be important to extend the study carried 
on in this paper to a larger set of GRB DLAs, in order to have an insight into 
regions of the universe complementary to those probed by QSO DLAs and by other types of astrophysical objects, detected 
by means of different techniques.

\acknowledgments
FC acknowledges financial contribution from contract ASI-INAF I/016/07/0.\\
J. X. P. is partially supported by NASA/Swift grants
NNG06GJ07G and NNX07AE94G and an NSF CAREER grant (AST-0548180).


\begin{thebibliography}{}
\bibitem[Berger et al. (2007)]{BER07} Berger, E.; Fox, D. B.; Kulkarni, S. R.; Frail, D. A.; Djorgovski, S. G., 2007, ApJ, 660, 504
\bibitem[Bloom et al. (1998)]{BLO98} Bloom, J. S.; Djorgovski, S. G.; Kulkarni, S. R.; Frail, D. A., 1998, ApJ, 507, L25
\bibitem[Bradamante et al. (1998)]{BRA98} Bradamante, F., Matteucci, F., D'Ercole, A. 1998, A\&A,337, 338
\bibitem[Calura \& Matteucci (2003)]{CAL03a} Calura, F., Matteucci, F., 2003,  ApJ, 596, 734
\bibitem[Calura \& Matteucci (2004)]{CAL04} Calura F., Matteucci F., 2004, MNRAS, 350, 351
\bibitem[Calura \& Matteucci (2006)]{CAL06} Calura F., Matteucci F., 2006, ApJ, 652, 889
\bibitem[Calura et al. (2003)]{CAL03b} Calura, F., Matteucci, F., \& Vladilo, G. 2003, MNRAS, 340, 59
\bibitem[Calura et al. (2008)]{CAL08} Calura, F.; Pipino, A.; Matteucci, F.,  2008, A\&A, 479, 669
\bibitem[Castro Cer\'on et al. (2006)]{CAS06} Castro Cer\'on, J. M.; Michałowski, M. J.; Hjorth, J.; Watson, D.; Fynbo, J. P. U.; Gorosabel, J., 2006, ApJ, 653, L85
\bibitem[Castro Cer\'on et al. (2008)]{CAS08} Castro Cer\'on, J. M.; Michałowski, M. J.; Hjorth, J.; Watson, D.; Fynbo, J. P., 2008 ApJ, in press, arXiv:0803.2235 
\bibitem[Chen et al. (2005)]{CHE05} Chen, H.-W., et al. 2005, ApJ, 634, L25
\bibitem[Chiappini et al. (2001)]{CHI01} Chiappini, C., Matteucci, F., Romano, D., 2001, ApJ, 554, 1044
\bibitem[Chiappini et al. (2003)]{CHI03} Chiappini, C., Matteucci, F., Meynet, G., 2003, A\&A, 410 257
\bibitem[Chiappini et al. (2005)]{CHI05} Chiappini, C., Matteucci, F., Ballero, S. K., 2005, A\&A, 437, 429
\bibitem[Chiappini et al. (2006)]{CHI06} Chiappini, C.; Hirschi, R.; Meynet, G.; Ekström, S.; Maeder, A.; Matteucci, F., 2006, A\&A, 449, L27
\bibitem[Christensen et al. (2004)]{CHR04} Christensen, L.; Hjorth, J.; Gorosabel, J., 2004, A\&A, 425, 913
\bibitem[Clayton et al. (1996)]{CLA96} Clayton, Geoffrey C., Green, J., Wolff, Michael J., Zellner, Nicolle E. B., Code, A. D., Davidsen, Arthur F., WUPPE Science Team, HUT Science Team, 1996, ApJ, 460 313 
\bibitem[D'Elia et al. (2007)]{DEL07} D'Elia, V., et al., 2007, A\&A, 467, 629
\bibitem[Dessauges-Zavadsky et al. (2002)]{DES02} Dessauges-Zavadsky, M., Prochaska, J. X.; D'Odorico, S.; 2002, A\&A, 391, 801
\bibitem[Dessauges-Zavadsky et al. (2004)]{DES04} Dessauges-Zavadsky, M.; Calura, F.; Prochaska, J. X.; D'Odorico, S.; Matteucci, F., 2004, A\&A, 416, 79
\bibitem[Dessauges-Zavadsky et al. (2006)]{DES06} Dessauges-Zavadsky, M.; Prochaska, J. X.; D'Odorico, S.; Calura, F.; Matteucci, F., 2006, A\&A, 445, 93
\bibitem[Dessauges-Zavadsky et al. (2007)]{DES07} Dessauges-Zavadsky, M.; Calura, F.; Prochaska, J. X.; D'Odorico, S.; Matteucci, F., 2007, A\&A, 470, 431 (D07)
\bibitem[Djorgovski et al. (2003)]{DJO03} Djorgovski, S. G.; Bloom, J. S.; Kulkarni, S. R., 2003, ApJ, 591, L13
\bibitem[Dwek (1998)]{DWE98} Dwek E., 1998, ApJ, 501, 643 (D98)
\bibitem[Dwek et al. (2007)]{DWE07} Dwek, E.; Galliano, F.; Jones, A. P., 2007, ApJ, 662, 927
\bibitem[Edmunds (2001)]{EDM01} Edmunds, M. G., 2001, MNRAS, 328, 223
%\bibitem[Fran\c cois et al. (2004)]{FRA04} Fran\c cois P., Matteucci F., Cayrel R., Spite M., Spite F., Chiappini C., 2004, A\&A, 421, 613
\bibitem[Fruchter et al. (1999)]{FRU99} Fruchter, Andrew S.et al., 1999, ApJ, 519, L13
\bibitem[Fynbo et al. (2006)]{FYN06} Fynbo, J. P. U, et al., 2006, A\&A, 451, L47
\bibitem[Fynbo et al. (2007)]{FYN07} Fynbo, J. P. U, Hjorth, J.; Malesani, D.; Sollerman, J.; Watson, D.; Jakobsson, P; Gorosabel, J.; Jaunsen, A. O., 2007, to appear in the proceedings of the Eleventh Marcel Grossmann Meeting on General Relativity, eds. H. Kleinert, R. T. Jantzen \& R. Ruffini, World Scientific, Singapore, 2007, (arXiv:astro-ph/0703458)
\bibitem[Fynbo et al. (2008)]{FYN08} Fynbo, J. P. U.; Prochaska, J. X.; Sommer-Larsen, J.; Dessauges-Zavadsky, M.; Moller, P., 2008, ApJ, accepted, arXiv0801.3273
\bibitem[Grevesse et al. (2007)]{GRE07} Grevesse, N.; Asplund, M.; Sauval, A. J.,  2007, SSRv, 130, 105
\bibitem[Henry \& Prochaska (2007)]{HEN07} Henry, R. B. C.; Prochaska, Jason X., 2007, PASP, 119, 962
\bibitem[Hopkins \&  Beacom (2006)]{HOP06} Hopkins, A. M., Beacom, J. F., 2006, ApJ, 651,142
\bibitem[Jones et al. (1994)]{JON94} Jones, A. P., Tielens, A. G. G. M., Hollenbach, D. J., McKee, C. F., 1994, ApJ, 433, 797
\bibitem[Keller et al. (2002)]{KELL02} Keller, L. P.; Hony, S.; Bradley, J. P.; Molster, F. J.; Waters, L. B. F. M.; Bouwman, J.; de Koter, A.; Brownlee, D. E.; Flynn, G. J.; Henning, T.; Mutschke, H.,  2002, Nature, 417, 148
\bibitem[Kimura et al. (2003)]{KIM03} Kimura H., Mann I., Jessberger E. K., 2003, ApJ, 582, 846
\bibitem[Issa et al. (1990)]{ISS90} Issa M. R., MacLaren I., Wolfendale A. W., 1990, A\&A, 236, 237
\bibitem[Iwamoto et al. (1999)]{IWA99} Iwamoto, K.; Brachwitz, F.; Nomoto, K.; Kishimoto, N.; Umeda, H.; Hix, W. R.; Thielemann, F.-K., 1999, ApJS, 125, 439
\bibitem[Inoue (2003)]{INO03} Inoue A. K., 2003, PASJ, 55, 901
\bibitem[Lanfranchi \& Matteucci (2003)]{LAN03} Lanfranchi, G., Matteucci, F., 2003, MNRAS, 345, 71 
\bibitem[Le Floc'h et al. (2002)]{LEF02} Le Floc'h, E.; Duc, P.-A.; Mirabel, I. F.; Sanders, D. B.; Bosch, G.; Rodrigues, I.; Courvoisier, T. J.-L.; Mereghetti, S.; Melnick, J., 2002, ApJ, 581, L81
\bibitem[Le Floc'h et al. (2006)]{LEF06} Le Floc'h, E.; Charmandaris, V.; Forrest, W. J.; Mirabel, I. F.; Armus, L.; Devost, D., 2006, ApJ, 642, 636
\bibitem[Levesque \& Kewley (2007)]{LEV07} Levesque, E. M.; Kewley, L. J., 2007, ApJ, 667, L121
\bibitem[Lisenfeld \& Ferrara (1998)]{LIS98} Lisenfeld U., Ferrara A., 1998, ApJ, 496, 145
\bibitem[Lodders (2003)]{LOD03} Lodders K., 2003, ApJ, 591, 1220
\bibitem[MacFadyen \&\ Woosley (1999)]{MAC99} MacFadyen A. and Woosley S., 1999, ApJ, 524, 262
\bibitem[Matteucci (1992)]{MAT92} Matteucci, F., 1992, ApJ, 397, 32
\bibitem[Matteucci (2001)]{MAT01} Matteucci, F., 2001, \emph{The chemical evolution of the Galaxy}, Astrophysics and space science library, Volume 253, Dordrecht: Kluwer Academic Publishers
\bibitem[Meynet \& Maeder (2000)]{MEY00} Meynet, G.,  Maeder, A. 2000, A\&A, 361, 101 
\bibitem[Meynet \& Maeder (2002)]{MEY02}Meynet, G.,  Maeder, A. 2002, A\&A, 390, 561 
\bibitem[Morgan \& Edmunds (2003)]{MOR03} Morgan H. L., Edmunds M. G., 2003, MNRAS, 343, 427
\bibitem[Matteucci (1986)]{MAT86} Matteucci F., 1986, MNRAS, 221, 911
%\bibitem[Matteucci \& Greggio (1986)]{MAT86} Matteucci F., Greggio L., 1986, A\&A, 154, 279
\bibitem[McKee (1989)]{MCK89} McKee C. F., 1989, in Allamandola L. J., Tielens A. G. G. M., eds, Interstellar Dust, Proc. IAU Symposium 135. Kluwer, Dordrecht, p. 431
\bibitem[Meynet \& Maeder (2002)]{MEY02} Meynet, G., Maeder, A., 2002, A\&A, 381, L25
\bibitem[Papovich et al. (2006)]{PAP06} Papovich, C., et al., 2006, in "Extreme Starbursts: Near and Far", editors: Yu Gao \& D. B. Sanders, in press, astro-ph/0601408
\bibitem[Penprase et al. (2006)]{PEN06} Penprase, B. E.; Berger, E.; Fox, D. B.; Kulkarni, S. R.; Kadish, S.; Kerber, L.; Ofek, E.; Kasliwal, M.; Hill, G.; Schaefer, B.; Reed, M.,  2006, ApJ, 646, 358
\bibitem[Pettini et al. (2008)]{PETT08} Pettini, M.; Zych, B. J.; Steidel, C. C.; Chaffee, F. H., 2008, MNRAS, 385, 2011
\bibitem[Phillips (2007)]{PHI07} Phillips, J. P., 2007, MNRAS, 381, 117
%\bibitem[Pipino et al. (2005)]{PIP05} Pipino A., Kawata D., Gibson B. K., Matteucci F., 2005, A\&A, 434, 553	
\bibitem[Price et al. (2007)]{PRI07} Price, P. A., et al., 2007, ApJ, 663, L57
\bibitem[Prochaska et al. (2004)]{PRO04}Prochaska, J. X.; Bloom, J. S.; Chen, H.-W.; Hurley, K. C.; Melbourne, J.; Dressler, A.; Graham, J. R.; Osip, D. J.; Vacca, W. D., 2004, ApJ, 611, 200
\bibitem[Prochaska et al. (2007a)]{PRO07a} Prochaska, J. X.; Chen, H.-W.; Bloom, J. S.; Dessauges-Zavadsky, M.; O'Meara, J. M.; Foley, R. J.; Bernstein, R.; Burles, S.; Dupree, A. K.; Falco, E.; Thompson, I. B., 2007, ApJS, 168, 231
\bibitem[Prochaska et al. (2007b)]{PRO07b} Prochaska, J. X.; Chen, H.-W.; Dessauges-Zavadsky, M.; Bloom, J. S., 2007, ApJ, 666, 267 (P07)
\bibitem[Ruffle et al. (1999)]{RUF99} Ruffle, D. P.; Hartquist, T. W.; Caselli, P.; Williams, D. A., 1999, MNRAS, 306, 691
\bibitem[Rubio et al. (2004)]{RUB04} Rubio, M.; Boulanger, F.; Rantakyro, F.; Contursi, A.,  2004, A\&A, 425, L1 
\bibitem[Savage \& Sembach (1996)]{SAV96} Savage B. D., Sembach K. R., 1996, ARA\&A, 34, 279
\bibitem[Savaglio et al.(2003)]{SAV03} Savaglio, S., Fall, S. M., Fiore, F.,  2003, ApJ, 585, 638
\bibitem[Savaglio et al.(2008)]{SAV08} Savaglio, S., Glazebrook, K., Le Borgne, D., 2008, ApJ, submitted , arXiv0803.271
\bibitem[Scappini et al. (2003)]{SCA03} Scappini, F.; Cecchi-Pestellini, C.; Smith, H.; Klemperer, W.; Dalgarno, A., 2003, MNRAS, 341, 657
\bibitem[Schmidt (1959)]{SCH59} Schmidt M., 1959, ApJ, 129, 243
\bibitem[Sembach et al. (2000)]{SEM00} Sembach, K. R.; Howk, J. C.; Ryans, R. S. I.; Keenan, F. P., 2000, ApJ, 528, 310
\bibitem[Spite et al. (2005)]{SPI05} Spite, M., et al.,  2005, A\&A, 430, 655
\bibitem[Starling et al. (2005)]{STA05} Starling, R. L. C., et al., 2005, A\&A, 442, L21
\bibitem[Sugerman et al. (2006)]{SUG06} Sugerman, B. E. K., et al., 2006, Science, 313, 196
\bibitem[Tumlinson et al. (2007)]{TUM07} Tumlinson, J., Prochaska, J. X., Chen, H.-W., Dessauges-Zavadsky, M., Bloom, J. S. 2007, ApJ, 668, 667
\bibitem[van den Hoek \& Groenwegen (1997)]{van97} van den Hoek L. B. \& Groenwegen M. A. T., 1997, A\&AS, 123, 305
\bibitem[Vladilo (2004)]{VLA04} Vladilo, G., 2004, A\&A, 421, 479
\bibitem[Vladilo et al. (2008)]{VLA08} Vladilo, G.; Prochaska, J. X.; Wolfe, A. M., 2008, A\&A, 478, 701
\bibitem[Vreeswijk et al. (2004)]{VRE04} Vreeswijk, P. M., et al., 2004, A\&A, 419, 927
\bibitem[Welty et al. (1997)]{WEL97}W elty, D. E., Lauroesch, J. T., Blades, J. C., Hobbs, L. M.,  York, D. G. 1997, ApJ, 489, 672 
\bibitem[Welty et al. (1999a)]{WEL99a} Welty, D. E., Hobbs, L. M., Lauroesch, J. T., Morton, D. C., Spitzer, L., York, D. G. 1999a, ApJS, 124, 465
\bibitem[Welty et al. (1999b)]{WEL99b}Welty, D. E., Frisch, P. C., Sonneborn, G., York, D. G. 1999b, ApJ, 512, 636
\bibitem[Welty et al. (2001)]{WEL01} Welty, D. E.; Lauroesch, J. T.; Blades, J. C.; Hobbs, L. M.; York, D. G.,  2001, ApJ, 554, L75
\bibitem[Whalen et al. (2008)]{WHA08} Whalen, D., Prochaska, J. X., Heger, A., Tumlinson, J., 2008, ApJ, submitted
\bibitem[Wiersema et al. (2007)]{WIE07} Wiersema, K., et al., 2007, A\&A, 464, 529
\bibitem[Wolf \& Podsiadlowski (2007)]{WOL07} Wolf, C.,  Podsiadlowski, P., 2007, MNRAS, 375, 1049
\bibitem[Woosley \& Weaver (1995)]{WOO95} Woosley, S.E., Weaver, T.A., 1995, ApJS, 101, 181
\bibitem[Zhukovska et al. (2008)]{ZHU08} Zhukovska S., Gail H.-P., Trieloff M., 2007, A\&A, 479, 453


\end{thebibliography}
\end{document}